\documentclass[usenatbib,usegraphicx]{mn2e}

\title[\spitzer\ mapping of pre-stellar cores]{The initial conditions of isolated star formation - VII. \spitzer\ mapping of pre-stellar cores}
\author[J.M.~Kirk, D.~Ward-Thompson and P.~Andr\'e]
{J.M. Kirk$^{1}$\thanks{Email: jason.kirk@astro.cf.ac.uk}, D. Ward-Thompson$^{1,2}$ and P. Andr\'e$^{3}$\\
$1$ School of Physics and Astronomy, Cardiff University, Queens Buildings, The Parade, Cardiff, CF24 3AA \\
$2$ Observatoire de Bordeaux, 2 rue de l'Observartoire, 33270, Floirac, France \\
$3$ CEA, DSM, DAPNIA, Service d'Astrophyique, C.E. Saclay, F-91191 Gif-sur-Yvette Cedex, France}

\newcommand\mnras{MNRAS}
\newcommand\aap{A\&A}
\newcommand\apj{ApJ}
\newcommand\apjs{ApJSS}
\newcommand\pasp{PASP}
\newcommand\nat{Nature}
\newcommand\skytel{Sky and Telesc.}

\newcommand\spitzer{{\it Spitzer}}
\newcommand\iso{{\it ISO}}

\newcommand\mum{$\mu$m}

\defcitealias{paper1}{Paper~I}
\defcitealias{paper2}{Paper~II}
\defcitealias{paper3}{Paper~III}
\defcitealias{paper4}{Paper~IV}
\defcitealias{paper5}{Paper~V}
\defcitealias{paper6}{Paper~VI}

\begin{document}

\maketitle
\begin{abstract}
We have retrieved \spitzer\ archive data of pre-stellar cores taken with the Multiband Imaging Photometer for \spitzer\ (MIPS) at a wavelength of 160 \mum. Seventeen images, containing eighteen cores, were constructed. Flux densities were measured for each core, and background estimates were made. Mean off-source backgrounds were found to be 48$\pm$10 MJy/sr in Taurus and 140$\pm$55 MJy/sr in Ophiuchus. Consistency was found between the MIPS 170\mum\ and ISOPHOT 160\mum\ calibration. Fourteen cores were detected both by MIPS and our previous submillimetre surveys. Spectral Energy Distributions (SEDs) were made for each core, using additional 24- and 70-\mum\ data from the {\it Spitzer} data archive, as well as previous infra-red and submillimetre data. Previous temperature estimates were refined, and new temperature estimates were made where no {\it Infrared Space Observatory} (\iso) data exist. A temperature range of 8--18K was found for the cores, with most lying in the range 10--13K. We discount recent claims that a large number of pre-stellar cores may have been misclassified and in fact contain low luminosity protostars detectable only by \spitzer. We find no new protostars in our sample other than that previously reported in L1521F. It is shown that this has a negligible effect on pre-stellar lifetime estimates.
\end{abstract}

\section{Introduction}

One of the greatest challenges currently facing star formation is the origin and nature of the cold dense molecular cloud cores where protostars form. Interstellar dust in these cores absorbs radiation at ultraviolet and visible wavelengths and re-radiates the energy in the far-infrared and submillimetre. This series of papers has used the great submillimetre and infrared telescopes to trace that emission and to explore the initial conditions of the star formation process.

\citet{1986beichman} compared the catalogue of dense cores surveyed in molecular lines by Myers and his collaborators \cite[see][and references therein]{1989bm} with the {\it IRAS} point source catalogue and identified a new class of dense molecular core. The lack of an infrared point source is taken as evidence that there is no embedded protostellar heat source in these `starless' cores. In \citetalias{paper1} \citep{paper1} of this series we termed those starless cores which we believed to be undergoing star formation `pre-protostellar' cores (subsequently shortened to `pre-stellar' for brevity). In \citetalias{paper2} and \citetalias{paper3} \citep*{paper2,paper3} we probed the 1.3m continuum emission from nine pre-stellar cores with the IRAM telescope.  

In \citetalias{paper4} \citep{paper4} a radiative transfer model fitted to C$^{18}$O observations of the core L1689B showed that there was freeze-out of CO towards the centre of the core, and that the combination of molecular line and continuum observations could constrain the temperature and density profiles of pre-stellar cores. The freeze out of molecules with density is widely established with, for example, similar results being found in the dense cores L1544 and B68 \citep{2002caselli,2006bergin}.

The submillimetre and millimetre spectral energy distribution (SED) of the observed pre-stellar cores from our first four papers showed that our observations were on the Rayleigh-Jeans side of a greybody spectrum that peaks somewhere around 100-200\mum. Low atmospheric transmission means that observations of this peak are normally impossible from the ground \citep[with the notable exception of THUMPER,][]{2005thumper}, and observers are forced to fall back on airborne \citep[e.g. Kuiper Airborne Observatory,][]{1976kuiper} or space based observatories. 

The {\it Infrared Space Observatory} (\iso) was a 65-cm diameter space telescope designed to observe between 2.5 and 240\mum\ \citep{1996iso}.  It was operated by the European Space Agency (ESA) and ceased operations in early 1998. We used the imaging photo-polarimeter (ISOPHOT) on \iso\ to search for the peak of the pre-stellar greybody spectrum by imaging eighteen cores at 90, 170 and 200\mum\ and presented our results in \citetalias{paper5} \citep*{paper5}. For each core maps of colour temperature were constructed using the pixel-to-pixel 170/200-\mum\ spectral index. These showed that most of the cores are either isothermal or are warmer at the edges, consistent with them lacking an internal heat source (i.e. a protostar) and with their surfaces being heated by the local interstellar radiation field. Full radiative transfer analysis also shows similar temperature gradients \citep{evans-cores2,2003sw}. 

Spectral energy distributions (SEDs) were constructed using the ISOPHOT data and our previous millimetre and submillimetre data. These were used to estimate the energy budget of the cores and showed that their infrared luminosity was comparable to the incident energy at shorter wavelengths. This confirms that the cores are indeed pre-stellar and are only heated by the local interstellar radiation field. Our 200 and 170\mum\ ISOPHOT observations showed the importance of far-infrared data to the accurate determination of the spectral energy distribution of pre-stellar cores. 

In \citetalias{paper6} \citep*{paper6} we used the Submillimetre Common User Bolometer Array (SCUBA) on the James Clerk Maxwell Telescope (JCMT) to conduct a survey towards 52 pre-stellar cores at 450 and 850-\mum\ (c.f. \citealt{2000serg,2006young}). Twenty-nine of the cores were detected and were separated into `bright' and `intermediate' groups based on their 850-\mum\ peak intensities. This split matches the list of evolved prestellar cores identified by \citet{2005evo} using N$_{2}$H$^{+}$ and N$_{2}$D$^{+}$ obervations. The ISO data from \citetalias{paper5} was used to calculate the physical parameters of the cores. This showed that while the normalised column density profiles of the cores have the same shape as an unstable Bonnor-Ebert Sphere \citep{1955ebert,1956bonnor} the implied physical characteristics of the spheres did not match those calculated for the cores. 

The Multiband Imaging Photometer for \spitzer\ \citep[MIPS,][]{2004mips} is a far-infrared camera on the {\it Spitzer Space Telescope} \citep{2004spitzer}. \spitzer\ is an 85-cm diameter cryogenically-cooled satellite telescope designed to operate from 3 to 160 \mum. MIPS provides \spitzer's long wavelength capabilities and has imaging wavebands at 24, 70, and 160 \mum\ with telescope-limited resolutions of 6, 18, and 40 arcsec respectively. The sensitivity, resolution and large field of view allows for the unparallelled rapid mapping of extended star formation regions. 

In this paper we present MIPS 160-\mum\ images and 24- \& 70-\mum\ flux densities of 18 regions around pre-stellar cores that we have downloaded from the \spitzer\ data archive. In Section \ref{obs} we outline our search strategy and post-pipeline processing. In Section \ref{results} we present and discuss the images and extract flux density measurements. In Section \ref{seds} we reprise our SEDs from \citetalias{paper5} and update them by including the new \spitzer\ data from this paper. In Section \ref{discuss} we discuss the implications for the nature and lifetimes of pre-stellar cores.

\section{Observations}\label{obs}

\subsection{Post-pipeline processing}

A search of the \spitzer\ data archive was made using the {\sc LEOPARD} software tool \citep{2005leopard}. The positions of pre-stellar cores from \citetalias{paper6} were cross-referenced against the archive and coincident 160-\mum\ observations were identified. The unfiltered basic calibrated data (BCD) version of the MIPS 160-\mum\ data were downloaded and re-gridded onto images with 15 arcsec diameter pixels using the {\sc MOPEX} routine {\sc MOSAIC} \citep{2005mopex}. Most of the archive data had been reduced with software pipeline version S11 while the region L1517 was reduced with pipeline version S13.2.

The MIPS 160-\mum\ array only has half the field-of-view of the larger 24 and 70-\mum\ arrays. For the slow and medium raster scanning modes this is not a significant problem as the field-of-view is stepped forward in increments smaller than the width of the 160-\mum\ array. However with the fast raster scanning mode, used to efficiently observe large areas, the array is stepped forward in increments equivalent to the width of the shorter wavelength arrays, resulting in the sky plane being under-sampled by a factor of 2 at 160-\mum. Thus the recommended best practise for fast-scanning is for 4 coverages over each point --- an offset forward and backscan over the same area to fully sample the sky and a complete repeat observation to reject spikes caused by cosmic ray strikes on the detectors \citep{2005mips}. 

Even with this redundancy it is necessary to reject outlying measurements during the re-gridding and to clean the fast-scanned maps afterwards. The {\sc KAPPA} task {\sc FFCLEAN} was used iteratively to reject pixels that lay more than 3$\sigma$ way from the local pixel mean and then the task {\sc FILLBAD} was used to fill the holes left by rejected pixels with a mean of the four-nearest good-pixels \citep{2004kappa}. The percentage of pixels filled by this method was found to be a product of scan-speed and absolute background intensity. The intensity of the good pixels was not altered by this method.

The medium scan maps did not require filtering while the fast scan areas required less that 30 per cent filling with all four-pixels for the mean coming from the eight neighbouring pixels. This gives a filtering scale of 3 pixels, 45 arcsec, comparable to the 40 arcsec MIPS beam at 160-\mum. Filtering in this manner will change the resolution, but only for those pixels actually infilled. We can estimate the magnitude of this change by adding the filtering length to the intrinsic resolution in the normal quadrature manner, but we weight the filtering scale by the percentage of pixels replaced. Using an upper-limit of 30 percent gives an estimate of the ``effective resolution'' of 47 arcsec, which is 3.1 pixels. 

Comparison of pixel statistics before and after filtering show that within the 50 per cent contour the the filtering did not significantly alter the mean pixel value, this is to be expected as the filling was based on the local mean. The peak intensity within that contour was also unaffected as, by definition, the infill value cannot be greater than that of any of the surrounding pixels. However, this does not preclude the true intensity peak being coincident with a blank pixel. The upper-limit of the number of pixels infilled is consistent with a 70 percent certainty of measuring the true peak intensity. The only core at variance with the above is L1689A which was close to the absolute saturation limit (see following section). Approximately 60 per cent of its area had to be reconstructed and it is possible that the peak intensity has been underestimated. This combination of {\sc FFCLEAN} and {\sc FILLBAD} was found to be relatively simple, robust and fast. 
	
\begin{table}
	\caption{\spitzer\ archive details. The Data Set Identifiers are given for all \spitzer\ archive data included in this paper.}
	\label{tab:datasets}
	
	\begin{tabular}{ll}
		\hline

		Region	& Data Set Identifier \\

		\hline
		L1498	&  12026624 \\
		L1521C \& SMM & 11229696/9952 \\
		L1521B	&  11229184/9440 \\
		L1521A \& D  & 11228672/8928 \\		
		L1521E \& F  & 11228160/8416	 \\
		L1517	&  12662784 \\ 
		L1512	&  12021248 \\
		L1696A	&  5759488, 5767680, 4321280/1536 \\
		L1709A	&  5758720, 5766912, 4321536/2048 \\
		L1689A	&  5758464/7952, 5766656/144, 4322048	 \\
		L1709C	&  5757952, 5766144, 4321792 \\
		L1689B	&  5749504, 5753867, 4321792  \\
		B68	&  12023552	 \\
		B133	&  12023808	 \\
		\hline

	\end{tabular}

\end{table}

Table \ref{tab:datasets} lists the `Data Set Identifier', a unique tracking number used to identify individual \spitzer\ observations, for each of our fields. The regions analysed in this paper were all taken with the MIPS fast-scanning mode, except for L1498, L1512, L1517, B68, and B133, which were taken with the medium-scanning mode. The Ophiuchus regions (L1689A/B, L1709A/C, and L1696A) are part of the area mapped by the `Cores to Disks' (C2D) legacy survey \citep{2003c2d}, although the BCDs for our maps were downloaded and processed by us independently of the C2D data release of this region. However, cross-checks have shown that the two reconstructions are consistent.

\subsection{Detector Saturation}

The \spitzer\ 160-\mum\ detector is a 2$\times$20 pixel stressed Ge:Ga photoconductor array. During an integration the rate of voltage increase on a pixel is measured by non-destructively reading the array 8 times per second. The `slope' of voltage with time is proportional to the incident intensity and is fitted by the automated software pipeline using pixel-by-pixel least-squares fitting \citep{2005mips}. Responsivity changes caused by bright sources and cosmic ray strikes are tracked by simulator flashes every two minutes. The absolute calibration is checked against celestial standards at the start of each observing campaign \citep{2006mips}.

The voltage ramp from a sufficiently bright source will saturate the instrument electronics before the end of the integration cycle. The saturation speed at 160\mum\ is 20 MJy/sr in 10 s \citep{2006mips}. The fast-scanning method has a per point integration time of 3 s so will begin to saturate above $\sim$70 MJy/sr. The number of useful readouts, and therefore the goodness of the slope calibration, decreases with increasing source brightness until the source instantly saturates the detector and makes an estimation of its brightness impossible. This does not mean that usable data cannot be taken during saturation, but there must be a sufficient number of voltage readouts to characterise the slope.

\begin{figure}
	\centering{\includegraphics[width=84mm]{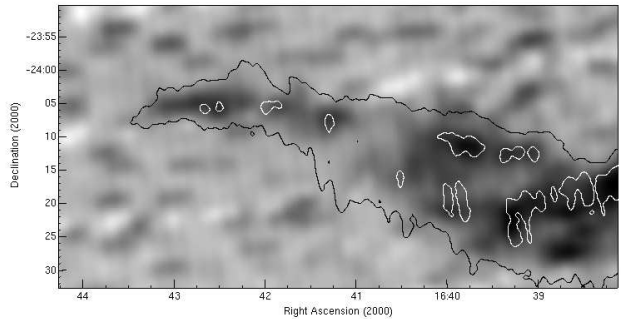}}
	\caption{Map of data coverage for a part of Ophiuchus. The greyscale shows the relative number of measurements that were propagated to the final map, black is low, white is high. The contours show intensity from the output map. The black contour is at 200 MJy/sr and the white contour is at 300 MJy/sr.}
	\label{fig:coverage}
\end{figure}

We can estimate the magnitude of the saturation problem by inspecting the number of slope measurements for which a valid intensity calibration could be made on a point-by-point basis. Figure \ref{fig:coverage} shows a map of the relative number of valid slope measurements superposed on contours of sky brightness. There is a distinct anti-correlation between source brightness and the relative number of valid measurements with the effect first becoming noticeable above 200 MJy/sr. The effect can be compensated for by increasing the number of observation repeats, but absolute saturation will still occur. From inspection of the data this absolute saturation appears to occur above 1000 MJy/sr. These intensities correspond to respective saturation times of 1 and 0.2 seconds --- the equivalent of $\sim$8 and $\sim$2-3 readouts -- with the brighter intensity saturating in the shorter time.

Of the cores included in this paper (see the following Section) only those from the Ophiuchus region regularly exceed 200 MJy/sr. L1709 A and C only just exceed this, while L1689B and L1696A double it. Only L1689A approaches the apparent level of absolute saturation. However, as we shall show in Section \ref{seds}, all their flux densities are consistent with those measured by ISOPHOT at 170\mum. Therefore we believe the data reduction and calibration holds good for all of the data presented in this paper.

\section{Results}\label{results}

\subsection{Source Positions and Flux Densities}

\begin{table*}
	\caption{\spitzer\ source positions, background intensities, and source flux densities. The quoted positions are the centroids of the 150-arcsec apertures used to measure the flux densities in Columns 7 and 8. Sources are listed in order of increasing Right Ascension. Column 4 lists the absolute peak intensity measured near the centroid before any background correction has been applied. Columns 5 and 6 list the mean and standard deviation of the source backgrounds as measured near to the minimum of the map (off-cloud) and close to the embedded core (on-cloud). Columns 7 and 8 list the background-subtracted flux densities in a 150 arcsec aperture centred on the centroid position using the backgrounds shown in Columns 5 and 6. The quantities in Columns 4--8 are quoted to 2 significant figures, the errors are quoted to the same number of decimal places, and the absolute calibration error is 20 per cent \citep{2006mips}. The error on the peak intensity (Column 4) is equal to the larger of the errors on the off-cloud or on-cloud background and is quoted to the same significance as the peak intensity. The error on the two backgrounds is the pixel-to-pixel standard deviation of a flat region, of size comparable to the flux density aperture, at the level of the background. Column 9 lists the SCUBA detection group from \citetalias{paper6} --- B is `bright', I is `intermediate' and n.d is `non-detection.'}
	\label{tab:positions}
	
	\begin{tabular}{lcccccccc}
		\hline

		Core	& Right Ascension & Declination & $I_{160}^{Peak}$ & \multicolumn{2}{c}{Background}  & \multicolumn{2}{c}{$S_{160}$} & SCUBA \\
			& (2000)	  & (2000)	&	& off-cloud & on-cloud	 & off-cloud & on-cloud	 & Group \\
			&		&		& [MJy/sr] & [MJy/sr] & [MJy/sr] & [Jy] & [Jy] \\

		\hline

		L1498	&	04$^{h}$ 10$^{m}$ 53.8$^{s}$   & $+$25\degr	09\arcmin 12\arcsec & 82$\pm$2	& 35$\pm$2	& 57$\pm$2 	& 16$\pm$1	& 6.8$\pm$0.1	& I \\
		L1521C	&	04$^{h}$ 19$^{m}$ 22.6$^{s}$   & $+$27\degr	14\arcmin 49\arcsec & 120$\pm$10& 49$\pm$3	& 83$\pm$3 	& 23$\pm$1	& 9.2$\pm$0.1	& n.d \\
		L1521SMM &	04$^{h}$ 21$^{m}$ 00.4$^{s}$   & $+$27\degr	02\arcmin 32\arcsec & $<$81	& 47$\pm$3	& 69$\pm$4 	& 3.2$\pm$0.1	& 0.86$\pm$0.10	& I \\
		L1521B	&	04$^{h}$ 24$^{m}$ 15.9$^{s}$   & $+$26\degr	37\arcmin 21\arcsec & 87$\pm$3	& 45$\pm$3	& 64$\pm$3 	& 14$\pm$1	& 5.9$\pm$0.2	& n.d. \\
		L1521A	& 	04$^{h}$ 26$^{m}$ 43.1$^{s}$   & $+$26\degr	16\arcmin 00\arcsec & $<$80	& 47$\pm$3 	& 71$\pm$3 	& 11$\pm$1	& 0.54$\pm$0.15	& n.d. \\
		                                                                                                    
		L1521D	& 	04$^{h}$ 27$^{m}$ 48.6$^{s}$   & $+$26\degr	18\arcmin 04\arcsec & 82$\pm$4	& 46$\pm$2	& 63$\pm$4 	& 10$\pm$1	& 3.4$\pm$0.2	& B	 \\                                                                         
		L1521F	&	04$^{h}$ 28$^{m}$ 37.9$^{s}$   & $+$26\degr	51\arcmin 08\arcsec & 140$\pm$10& 59$\pm$4	& 110$\pm$6 	& 25$\pm$1	& 5.6$\pm$0.3	& B	 \\                                                                        
		L1521E	&	04$^{h}$ 29$^{m}$ 15.0$^{s}$   & $+$26\degr	14\arcmin 10\arcsec & 94$\pm$4	& 39$\pm$2	& 78$\pm$4 	& 17$\pm$1	& 1.0$\pm$0.2	& I	 \\
		L1517A	&	04$^{h}$ 55$^{m}$ 17.0$^{s}$   & $+$30\degr	32\arcmin 53\arcsec & 110$\pm$10& 69$\pm$3	& 85$\pm$3 	& 15$\pm$1	& 8.9$\pm$0.1	& I \\ 
		L1517B	&	04$^{h}$ 55$^{m}$ 15.5$^{s}$   & $+$30\degr	37\arcmin 38\arcsec & 100$\pm$10& 69$\pm$3	& 85$\pm$3 	& 9.8$\pm$0.1	& 3.2$\pm$0.1	& B \\ 
		
		L1512	&	05$^{h}$ 04$^{m}$ 10.7$^{s}$   & $+$32\degr	42\arcmin 56\arcsec & 100$\pm$10& 49$\pm$3	& 68$\pm$2 	& 18$\pm$1	& 10$\pm$1	& I \\
		L1696A	&	16$^{h}$ 28$^{m}$ 29.3$^{s}$   & $-$24\degr	16\arcmin 51\arcsec & 470$\pm$40& 130$\pm$40	& 290$\pm$20 	& 110$\pm$10	& 45$\pm$1	& B \\
		L1709A	&	16$^{h}$ 30$^{m}$ 47.8$^{s}$   & $-$23\degr	41\arcmin 36\arcsec & 220$\pm$10& 140$\pm$10	& 180$\pm$10 	& 26$\pm$1	& 9.8$\pm$0.5	& I \\
		
		L1689A	&	16$^{h}$ 32$^{m}$ 15.8$^{s}$   & $-$25\degr	03\arcmin 40\arcsec & 910$\pm$30& 280$\pm$20	& 450$\pm$30 	& 200$\pm$10	& 130$\pm$10	& I \\
		L1709C	&	16$^{h}$ 33$^{m}$ 59.3$^{s}$   & $-$23\degr	40\arcmin 18\arcsec & 230$\pm$10& 95$\pm$9	& 140$\pm$10 	& 46$\pm$1	& 25$\pm$1	& n.d. \\
		L1689B	&	16$^{h}$ 34$^{m}$ 50.6$^{s}$   & $-$24\degr	38\arcmin 15\arcsec & 460$\pm$30& 160$\pm$20	& 290$\pm$30 	& 88$\pm$1	& 37$\pm$1	& B  \\
		
		B68	&	17$^{h}$ 22$^{m}$ 38.7$^{s}$   & $-$23\degr	50\arcmin 23\arcsec & 140$\pm$10& 73$\pm$4	& 96$\pm$5 	& 21$\pm$1	& 12$\pm$1	& I \\
		B133	&	19$^{h}$ 06$^{m}$ 09.8$^{s}$   & $-$06\degr	53\arcmin 04\arcsec & 160$\pm$10& 63$\pm$2	& 93$\pm$6 	& 32$\pm$1	& 19$\pm$1	& B	 \\

		\hline

	\end{tabular}

\end{table*}

Table \ref{tab:positions} lists the \spitzer\ positions of the SCUBA core regions from \citetalias{paper6} that we found had been observed by MIPS and stored in the \spitzer\ data archive. The seventeen regions that were downloaded contained nineteen separate SCUBA core search positions from \citetalias{paper6}. The SCUBA non-detection of  L1517D is confirmed by MIPS not to be a real core (see below), so hereafter we limit our discussion to the remaining 18 cores. For those 18 cores we created maps that matched the approximate area ($\sim$18 arcmin square) of our ISOPHOT observations from \citetalias{paper5}. These maps (see below) contain additional sources to those mapped with SCUBA, however, in this paper we restrict ourselves to an examination of the SCUBA selected sources and leave analysis of the other objects to a later paper. 

A 150-arcsec diameter aperture was placed at each 850-\mum\ position and a centroid was performed to determine the equivalent 160-\mum\ position --- these are the positions listed in Columns 2 and 3 of Table \ref{tab:positions}. In most cases the coordinates are coincident with the SCUBA positions to within a difference equivalent to the effective resolution of the MIPS maps (47 arcsec, see above). The maximum pixel value in the region of the centroid was taken to be the peak intensity of the core and is listed in Column 4. The background-subtracted (see below) flux densities measured from the centroid apertures are listed in Columns 7 and 8. The core L1521SMM has two peaks at 850\mum, but only one of these was detected at 160\mum. Given their close proximity we use half-sized apertures to avoid overlap and list positions determined from the SCUBA maps.

A major difference between the operation of MIPS on \spitzer\ and SCUBA on the JCMT is that the JCMT has a chopping secondary mirror which is used to reject bright atmospheric emission. This creates a problem when trying to compare data from both telescopes as \spitzer\ is sensitive to the entire dynamic range and spatial scale of the source emission, but SCUBA is only sensitive to the relative brightness and spatial scales within a region determined by the chop throw of the secondary mirror (see section 5.4 of \citetalias{paper6}). It is therefore necessary to subtract a background from each \spitzer\ image in order to directly compare observations from each instrument.

\begin{figure}
	\centering{
		\includegraphics[width=40mm,angle=270]{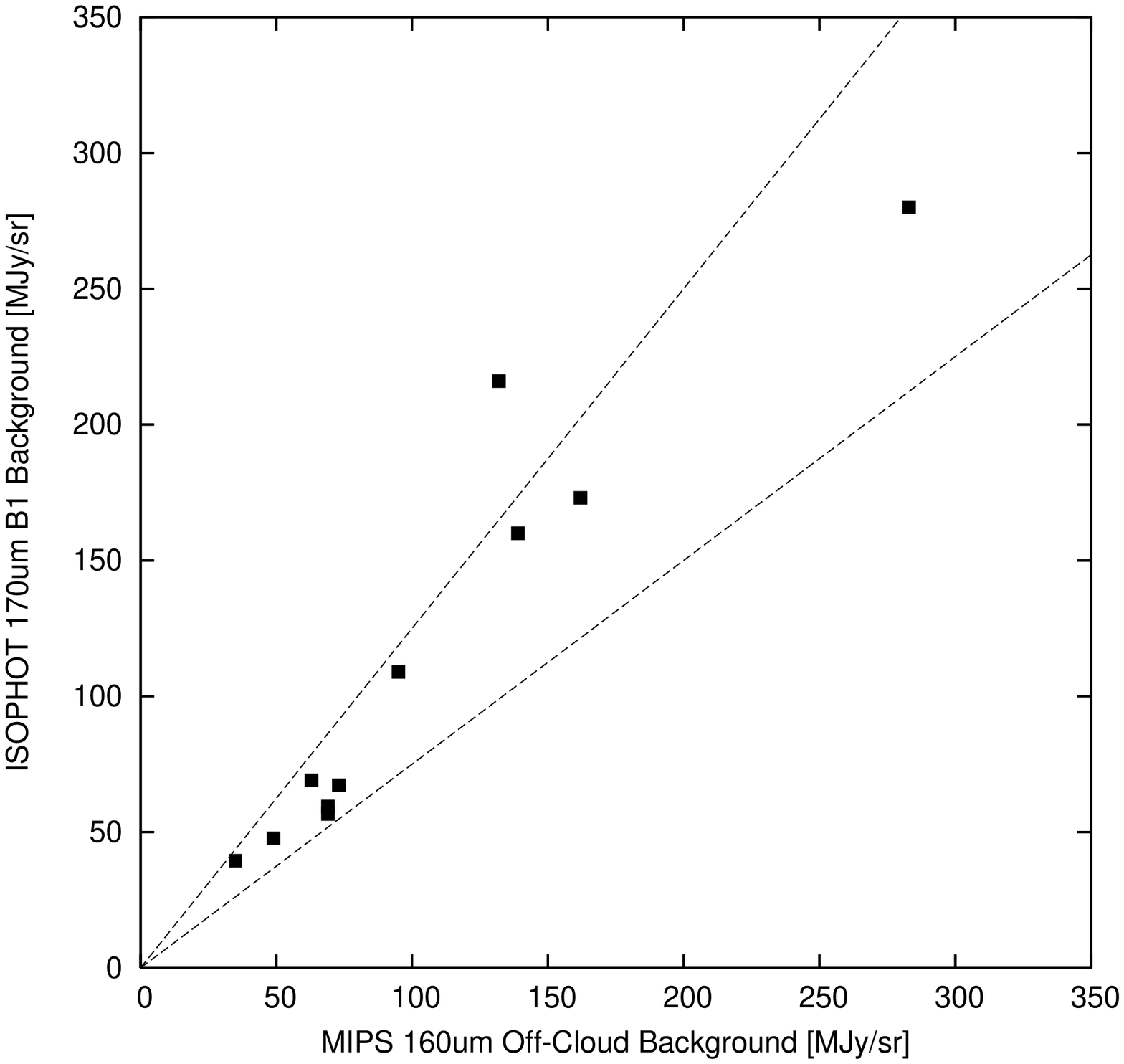}
		\includegraphics[width=40mm,angle=270]{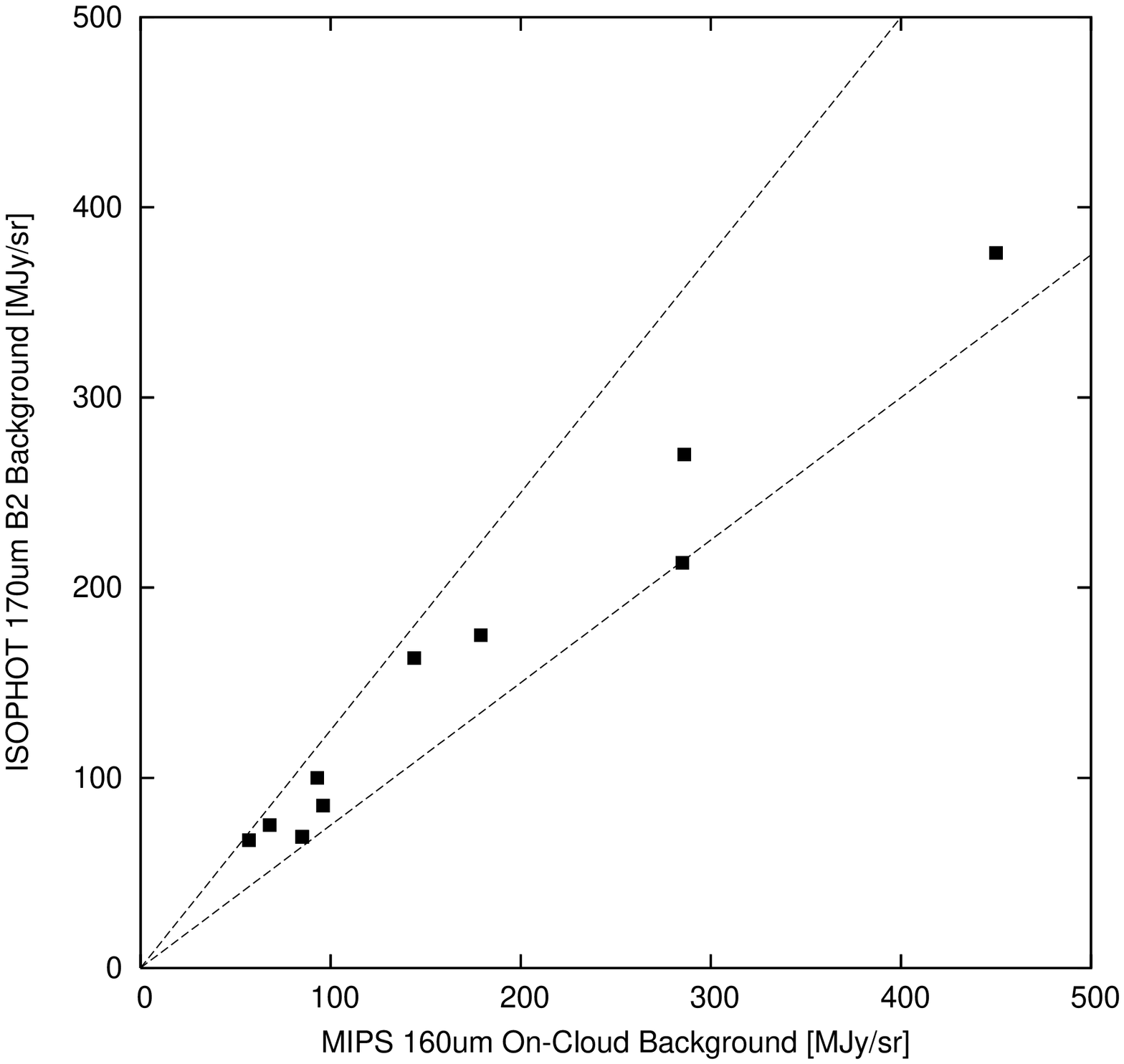}
	}
	\caption{A comparison of the off-cloud and on-cloud MIPS 160-\mum\ backgrounds with the equivalent ISOPHOT 170-\mum\ backgrounds from \citetalias{paper5}. The two fan-lines denote the $\pm$25 per cent envelope that would be expected from the 30 percent ISOPHOT and 20 percent MIPS calibration errors.}
	\label{fig:back}
\end{figure}

In the manner of our earlier ISOPHOT analysis we estimate background intensities from the images. In \citetalias{paper5} we estimated three different background levels from the ISOPHOT maps -- background 1 (B1) was the lowest pixel in the map, background 2 was estimated from the extended emission that surrounded the closed contours of the core, and a third ``best-fit'' background was estimated by fitting a slope to the same extended emission. MIPS and ISOPHOT have comparable beam sizes, but ISOPHOT undersampled its beam as a trade off for increased sensitivity. The increased information in the MIPS maps makes it harder to fit a simple 2D plane to the background so we forego this method in favour of the other two background subtraction methods from \citetalias{paper5}.  

In order to duplicate the first ISOPHOT background we constained our map sizes to the same approximate map size as the ISOPHOT observations. We then measured an ``off-cloud'' background across the lowest region of the map. The off-cloud background and standard deviation for each core is listed in column 5 of Table \ref{tab:positions} and is shown as a dotted contour on the maps in Figures \ref{fig:cores1} and \ref{fig:cores2}. The left panel of Figure \ref{fig:back} shows a graph of the ISOPHOT B1 background against the new off-cloud background. All points except one outlier (L1689A) are within the $\pm$25 per cent calibration error fan (based on a mean of the 30 per cent ISOPHOT and 20 per cent MIPS calibration errors). 

The agreement between backgrounds shows that both instruments have consistent calibrations, but to be able to compare our measured flux densities with chopped SCUBA data we need to subtract a greater background level. To allow this an ``on-cloud'' background was measured at the level of the extended or filamentary material which appears to surround the closed contours of each core. A further constraint on this background is that it has to be comparable in distribution to the size of the SCUBA chop throw (2 arcmin in this case) and in this aspect it is analagous to a 4 arcmin diameter sky annulus. The on-cloud background and standard deviation are listed in column 6 of Table \ref{tab:positions} and are shown as the lowest solid contour on the maps in Figures and \ref{fig:cores2}. The on-cloud background is comparable to the B2 ISOPHOT background; the right panel of Figure \ref{fig:back} shows a graph of the on-cloud background against the B2 background. As with the off-cloud case these two backgrounds are in agreement. 

The peak intensities listed in column 4 of Table \ref{tab:positions} are dominated by the intensity of the background with the on-cloud background contributing, on average, two-thirds to the full range of the peak intensity. The standard deviations of the two backgrounds are estimated from the pixel-to-pixel variation across the relatively flat region used to estimate each background.

The majority of the data in our sample comes from two distinct star formation regions --- the Taurus region at $\sim$04 hours R.A. and the Ophiuchus region at  $\sim$16 hours R.A. The mean off-cloud background showed a significant difference between these two regions. The weighted mean off-cloud background was 48$\pm$10 MJy/sr in Taurus and 140$\pm$55 MJy/sr in Ophiuchus. From \citetalias{paper5} the equivalent mean B1 backgrounds at 170-\mum\ were 58$\pm$8 MJy/sr in Taurus and 142$\pm$41 MJy/sr in Ophiuchus. This further shows that we are probing the same emission with each instrument and that their absolute calibrations are consistent. 

\subsection{Imaging Data}

\begin{figure*}
	\begin{tabular}{ccc}
		\vspace{2mm}
		\includegraphics[height=0.28\textwidth]{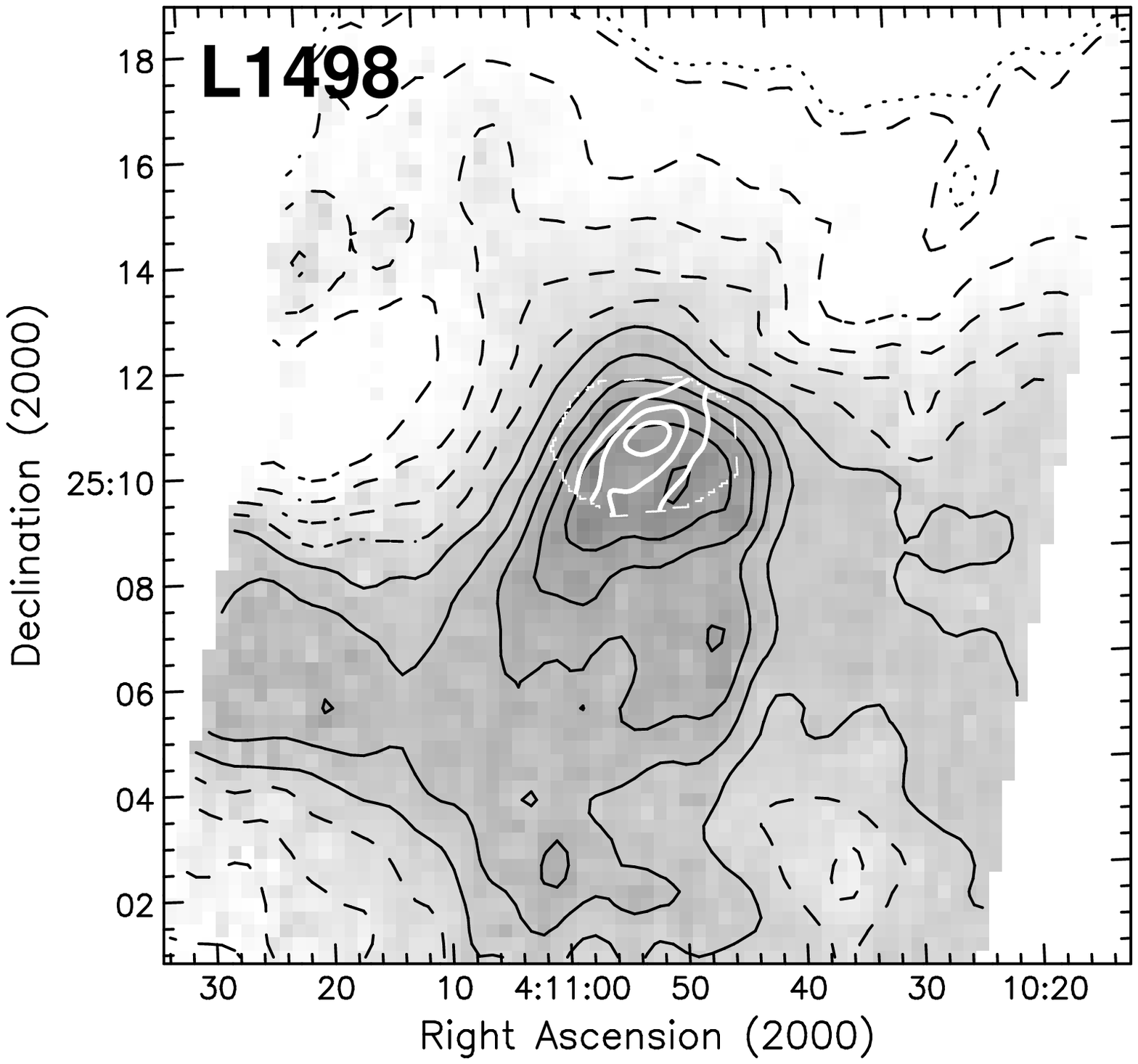} &
		\includegraphics[height=0.28\textwidth]{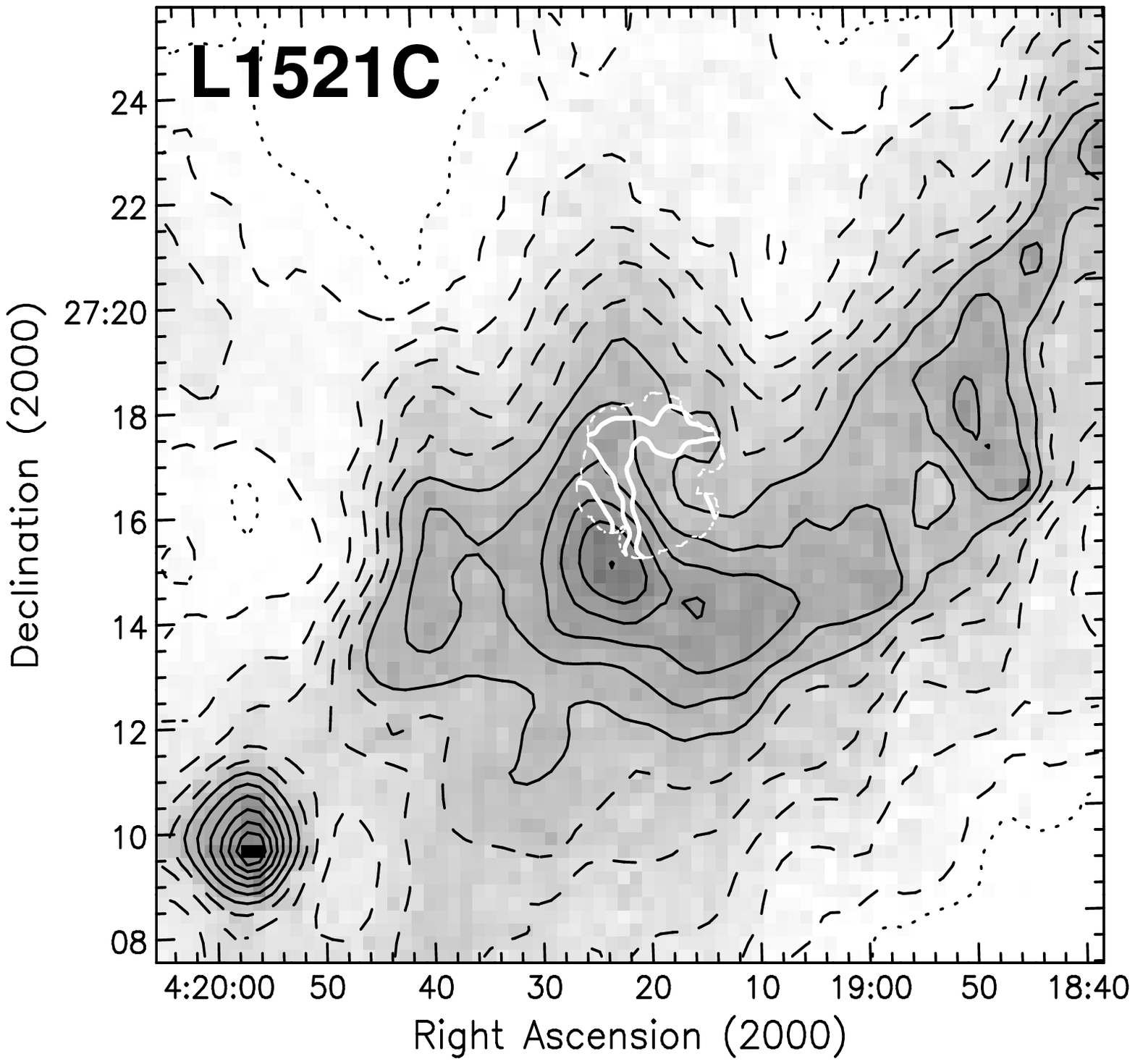} &
		\includegraphics[height=0.28\textwidth]{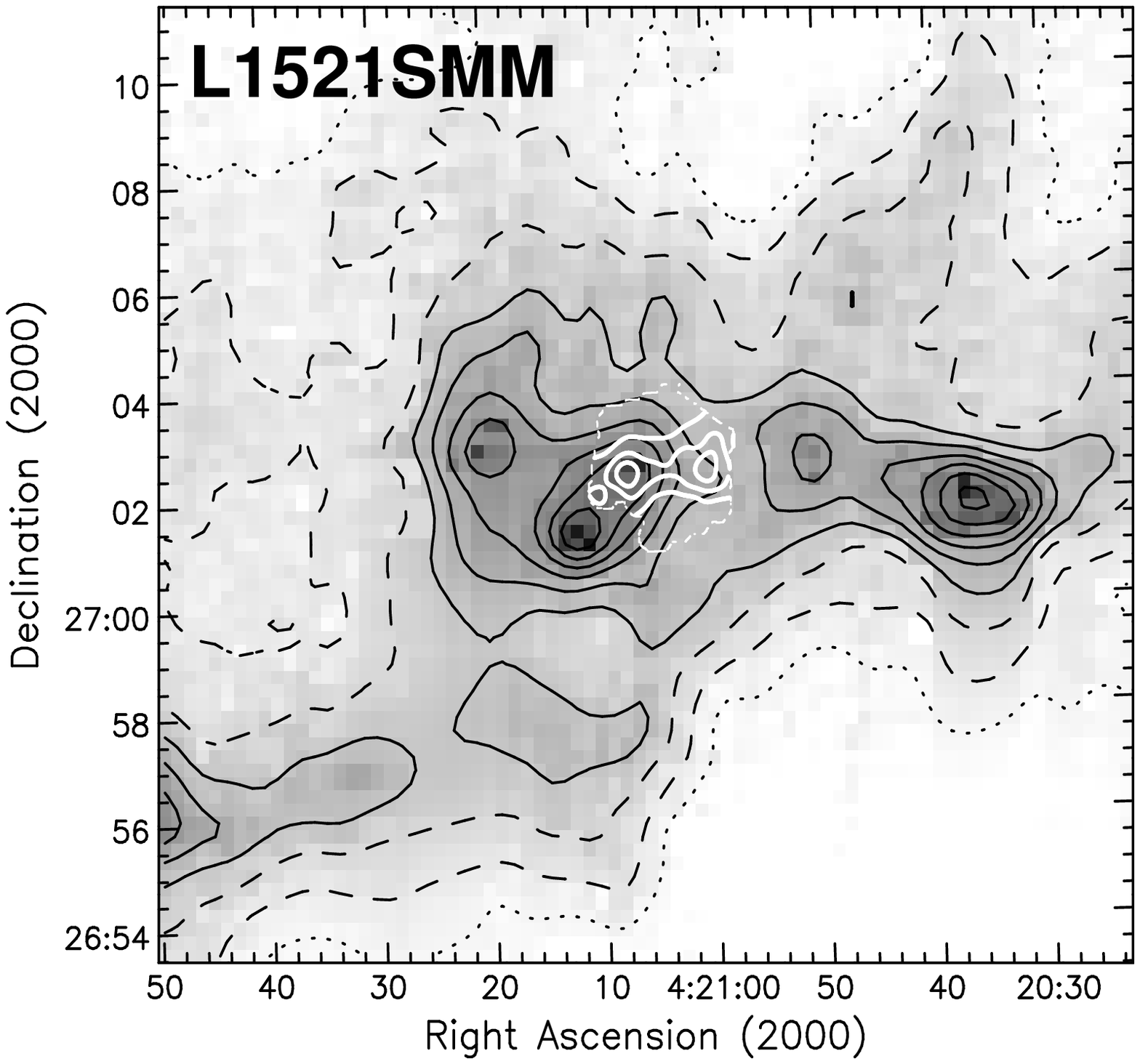} \\
		
		\vspace{2mm}
		\includegraphics[height=0.28\textwidth]{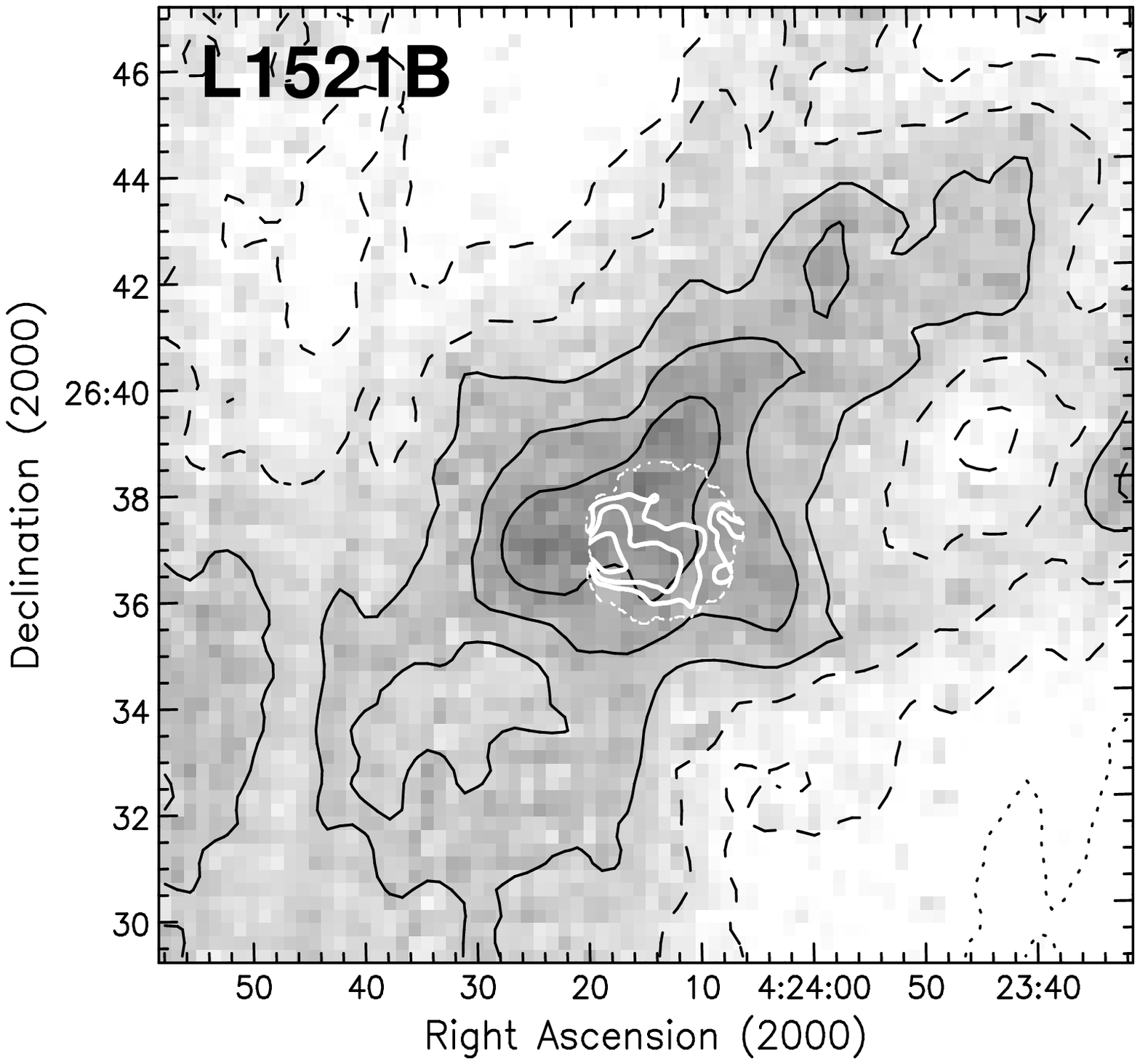} &
		\includegraphics[height=0.28\textwidth]{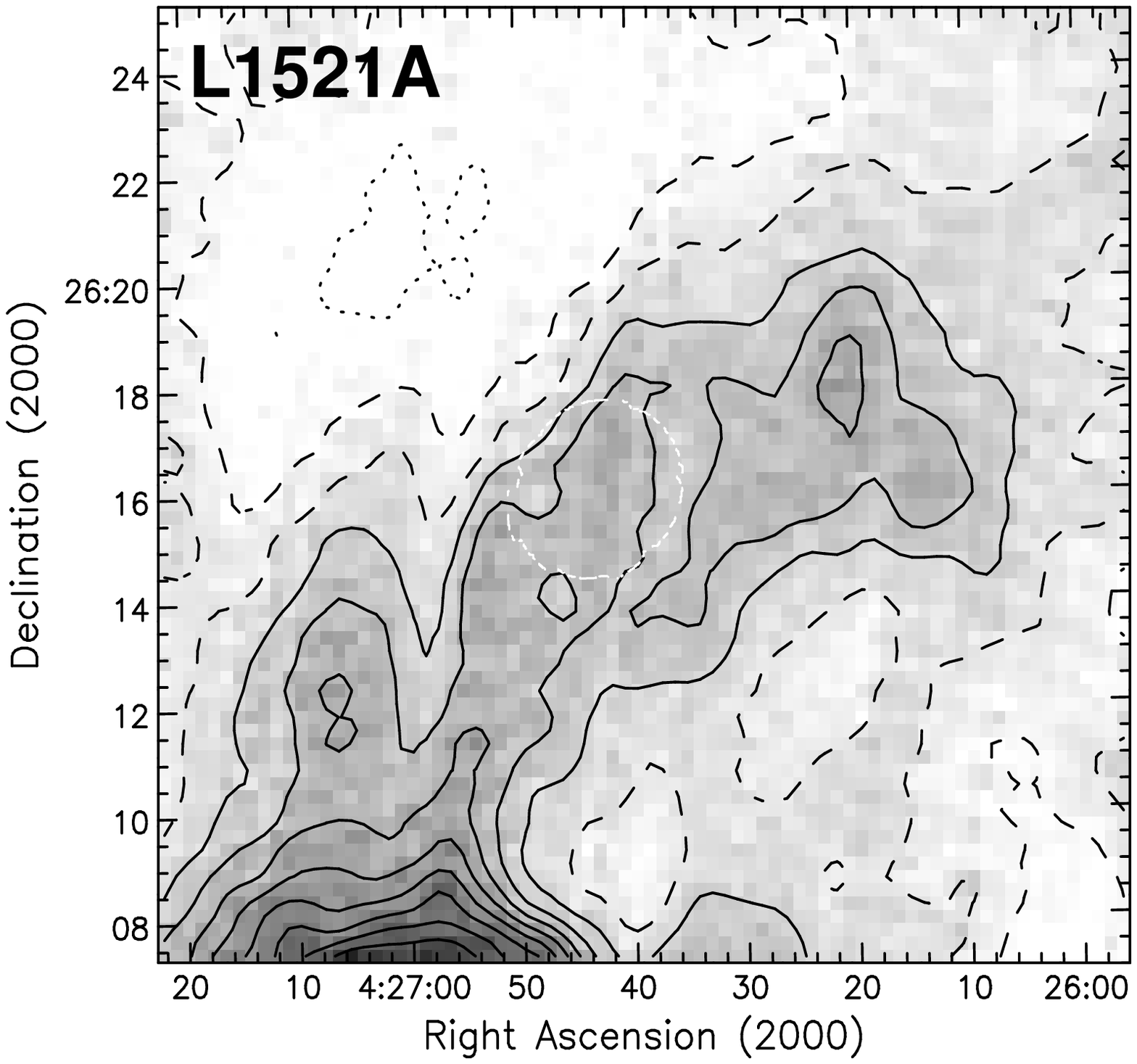} &
		\includegraphics[height=0.28\textwidth]{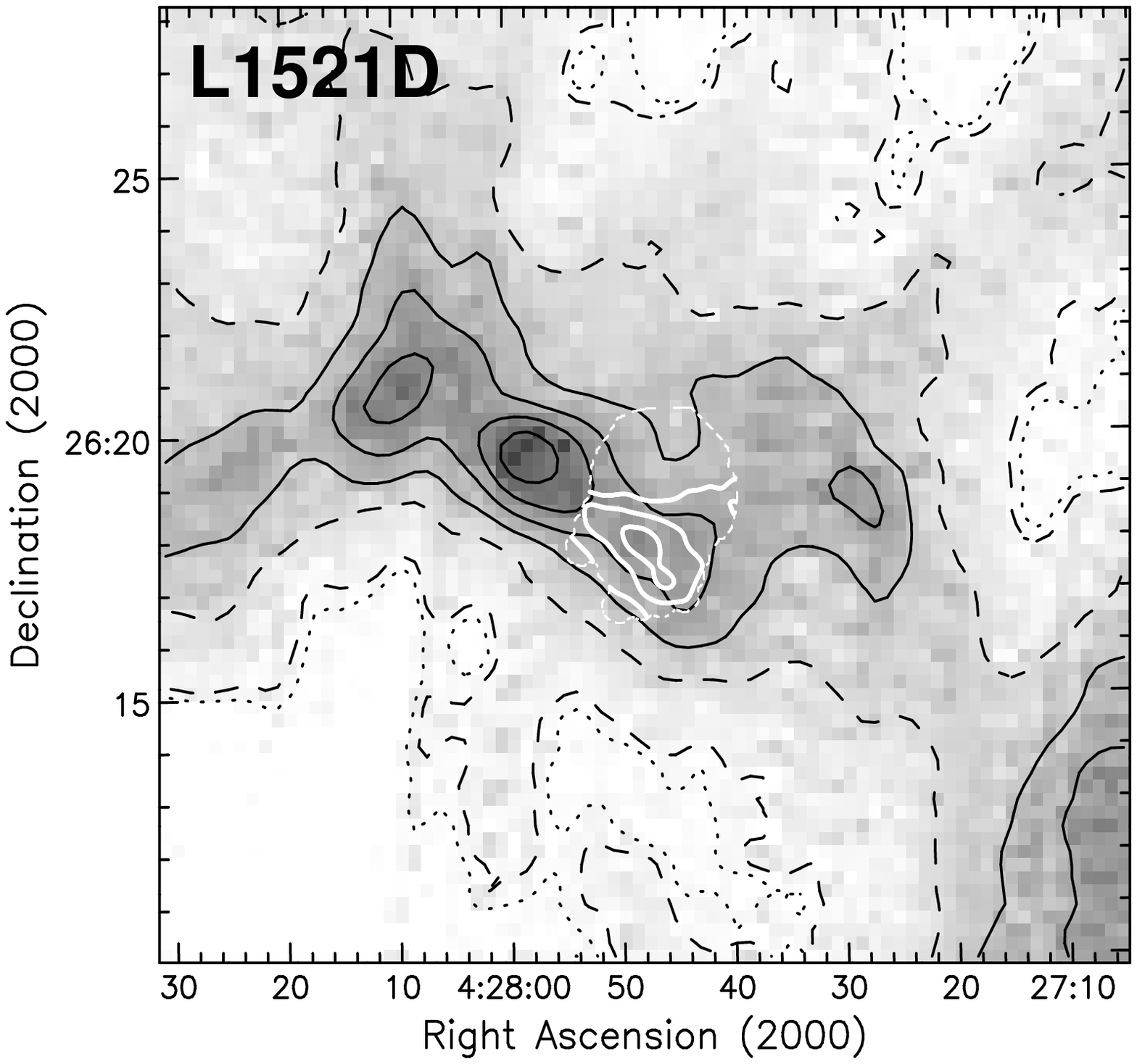} \\
		
		\vspace{2mm}
		\includegraphics[height=0.28\textwidth]{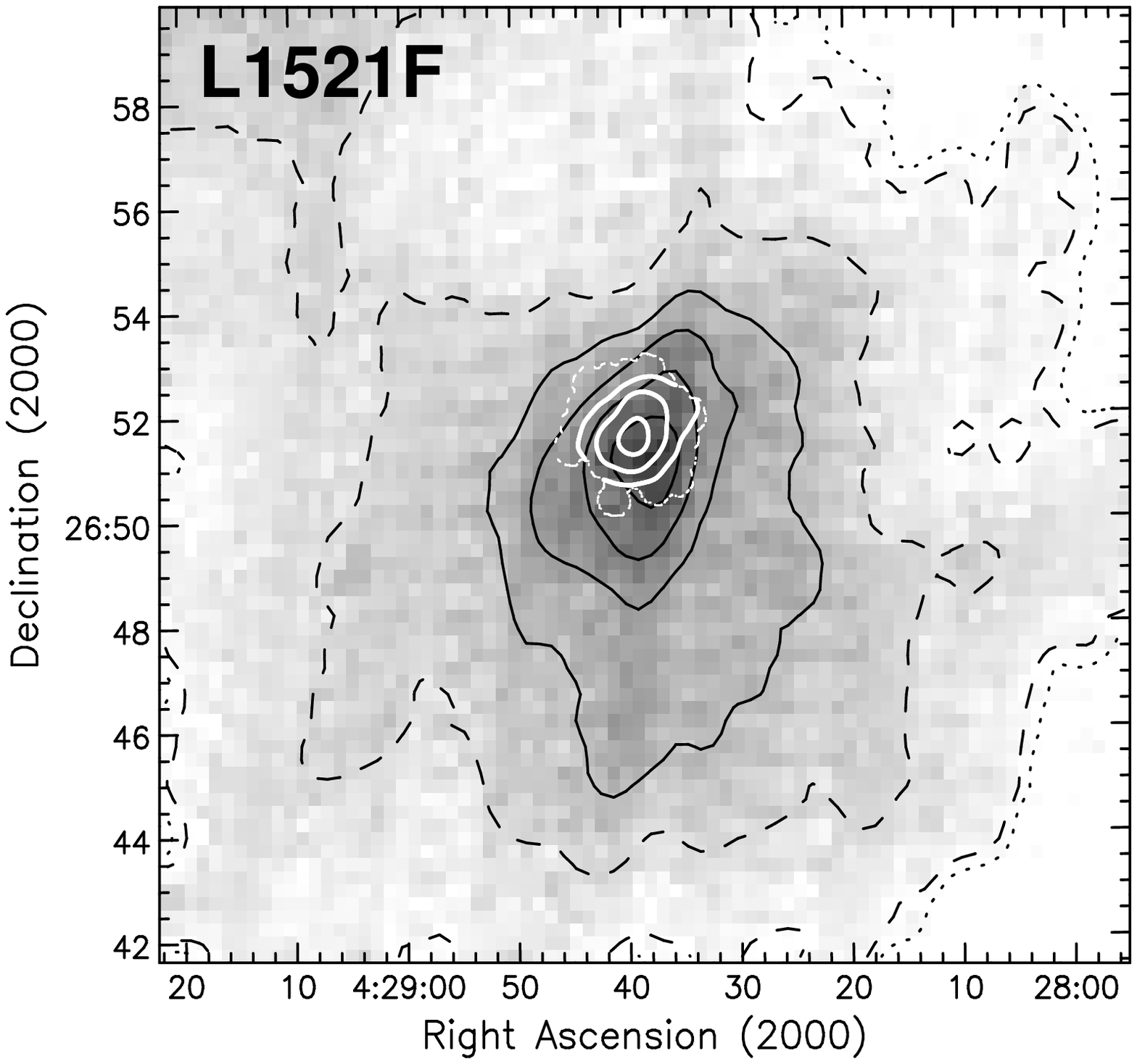} &
		\includegraphics[height=0.28\textwidth]{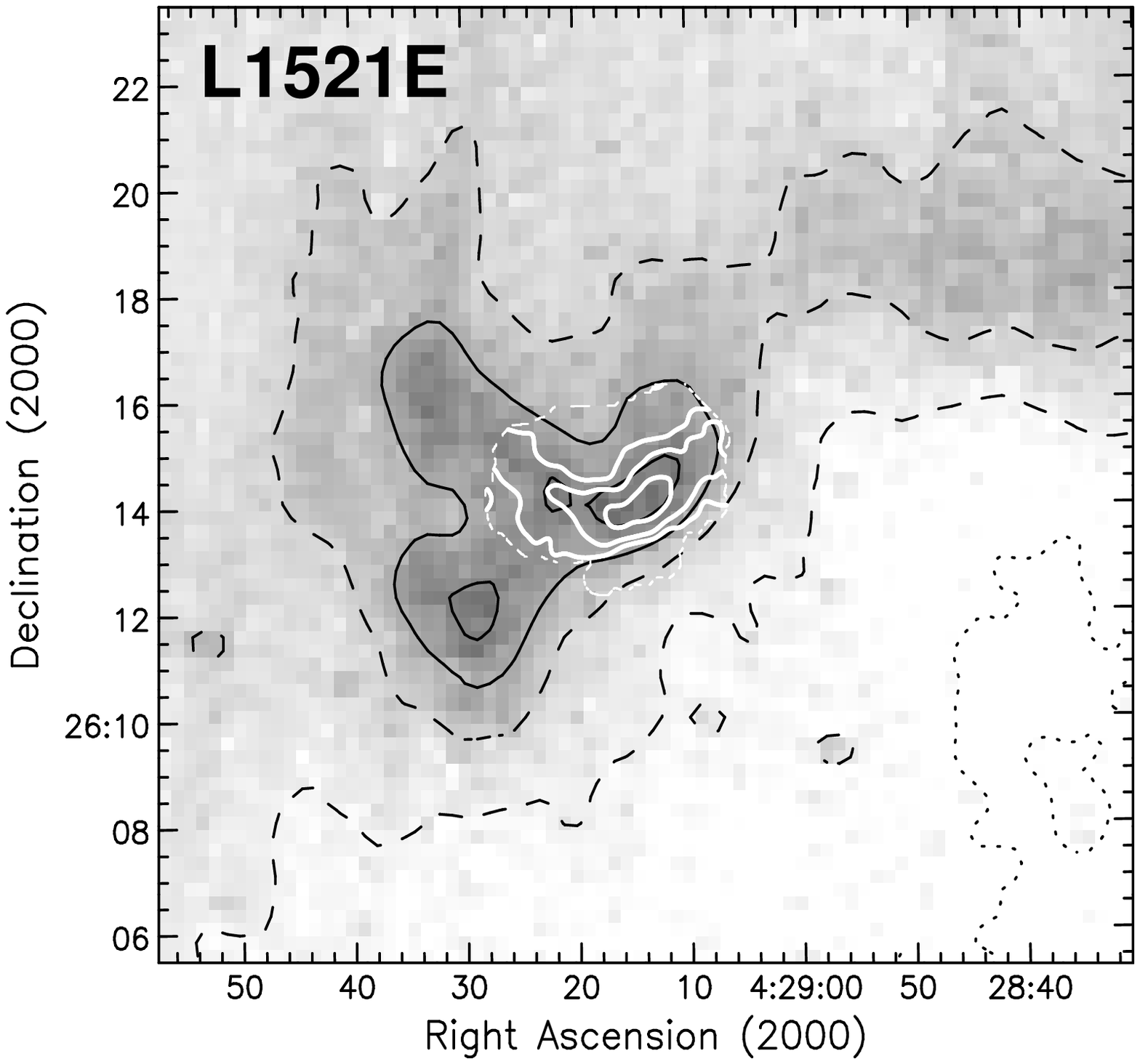} &
		\includegraphics[height=0.28\textwidth]{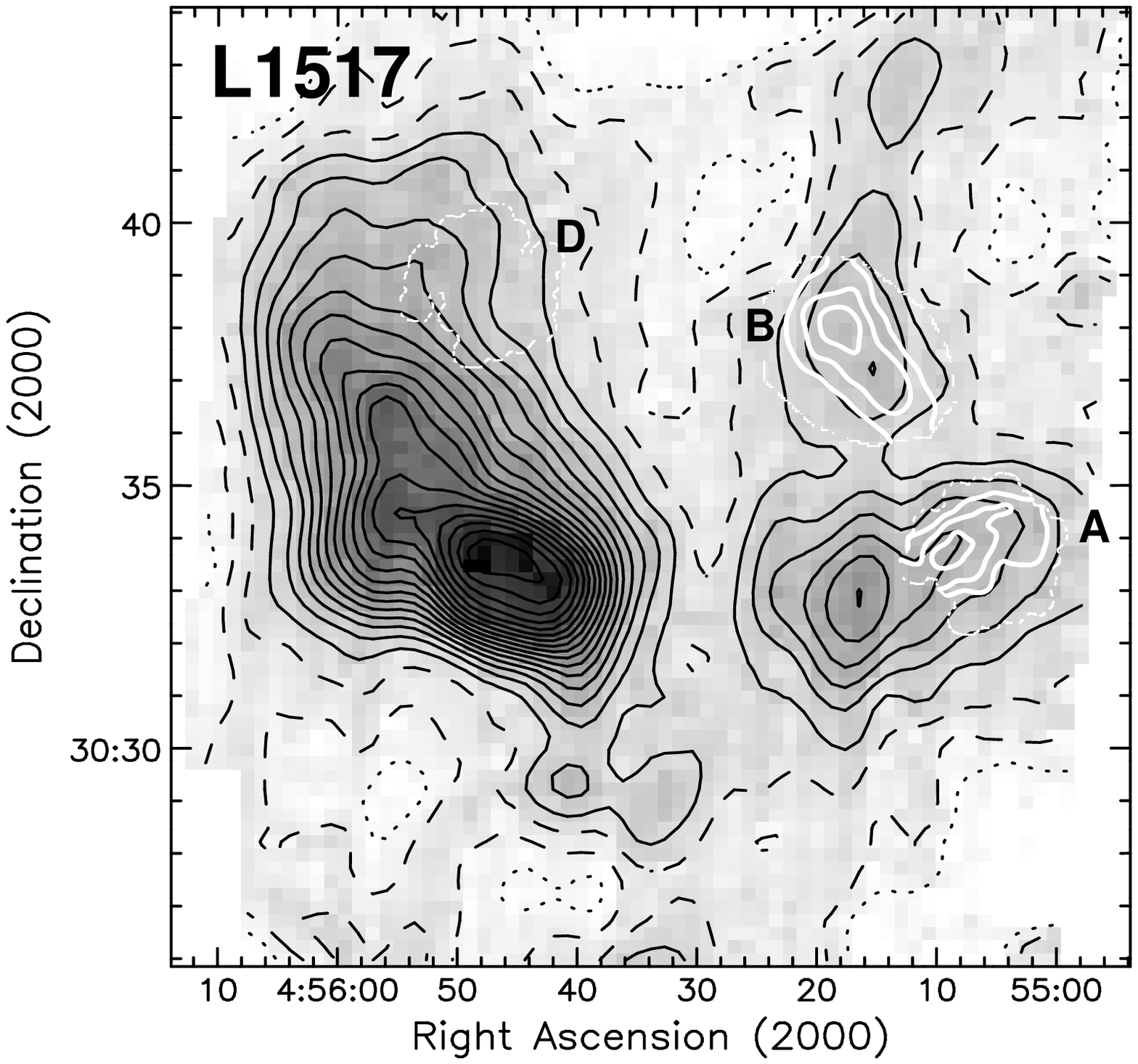} \\
		
		\includegraphics[height=0.28\textwidth]{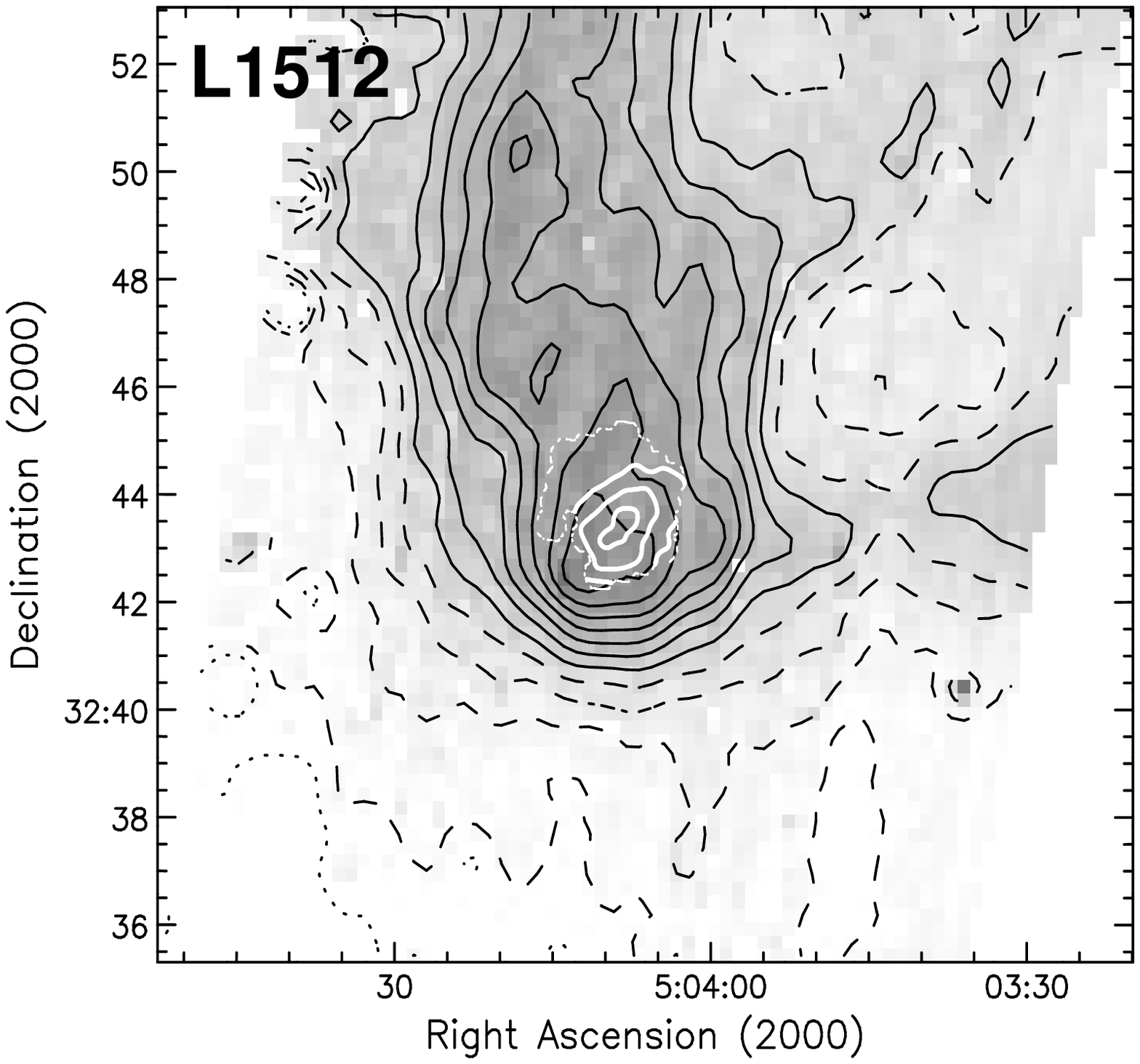} &
		\includegraphics[height=0.28\textwidth]{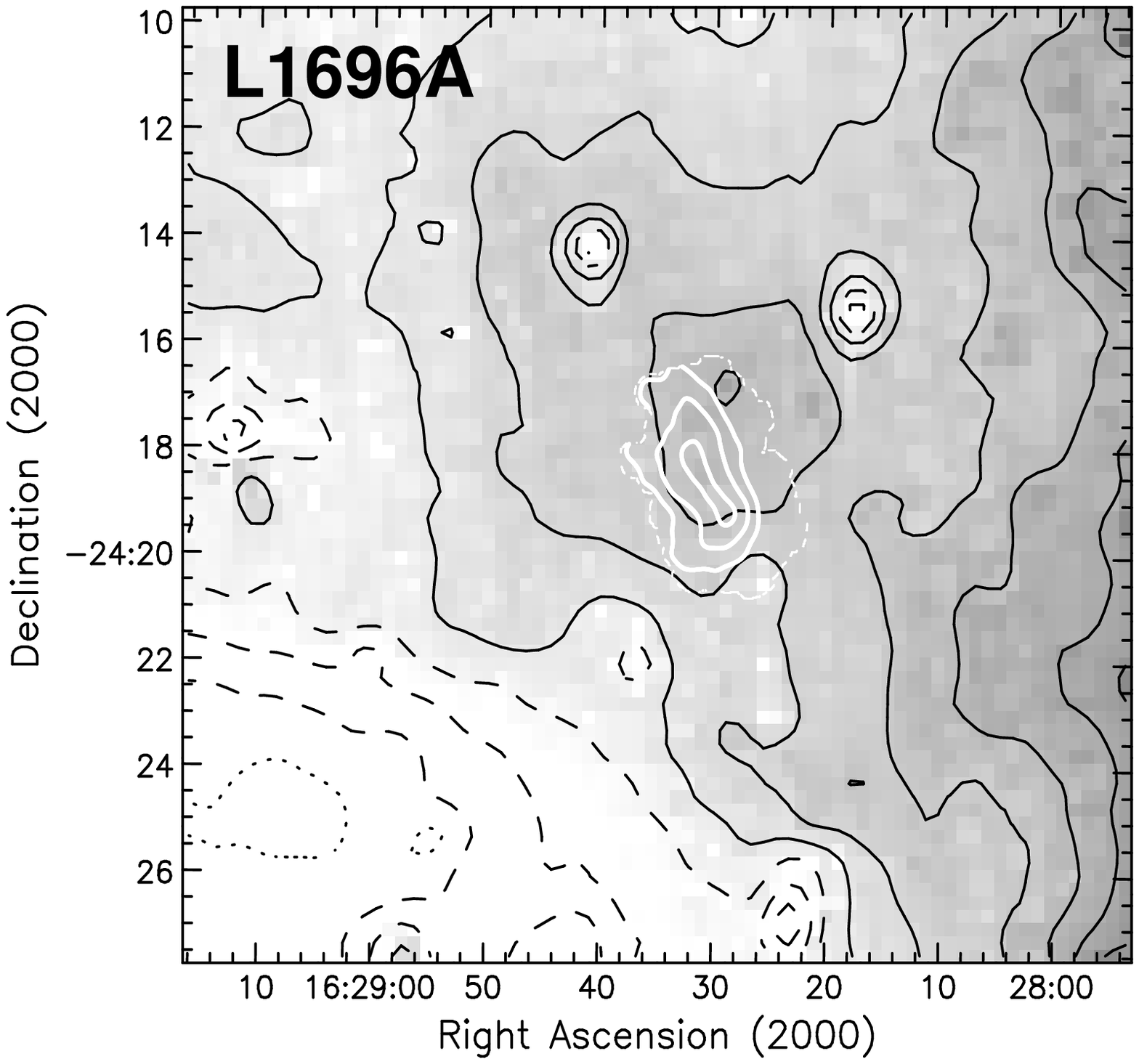} &
		\includegraphics[height=0.28\textwidth]{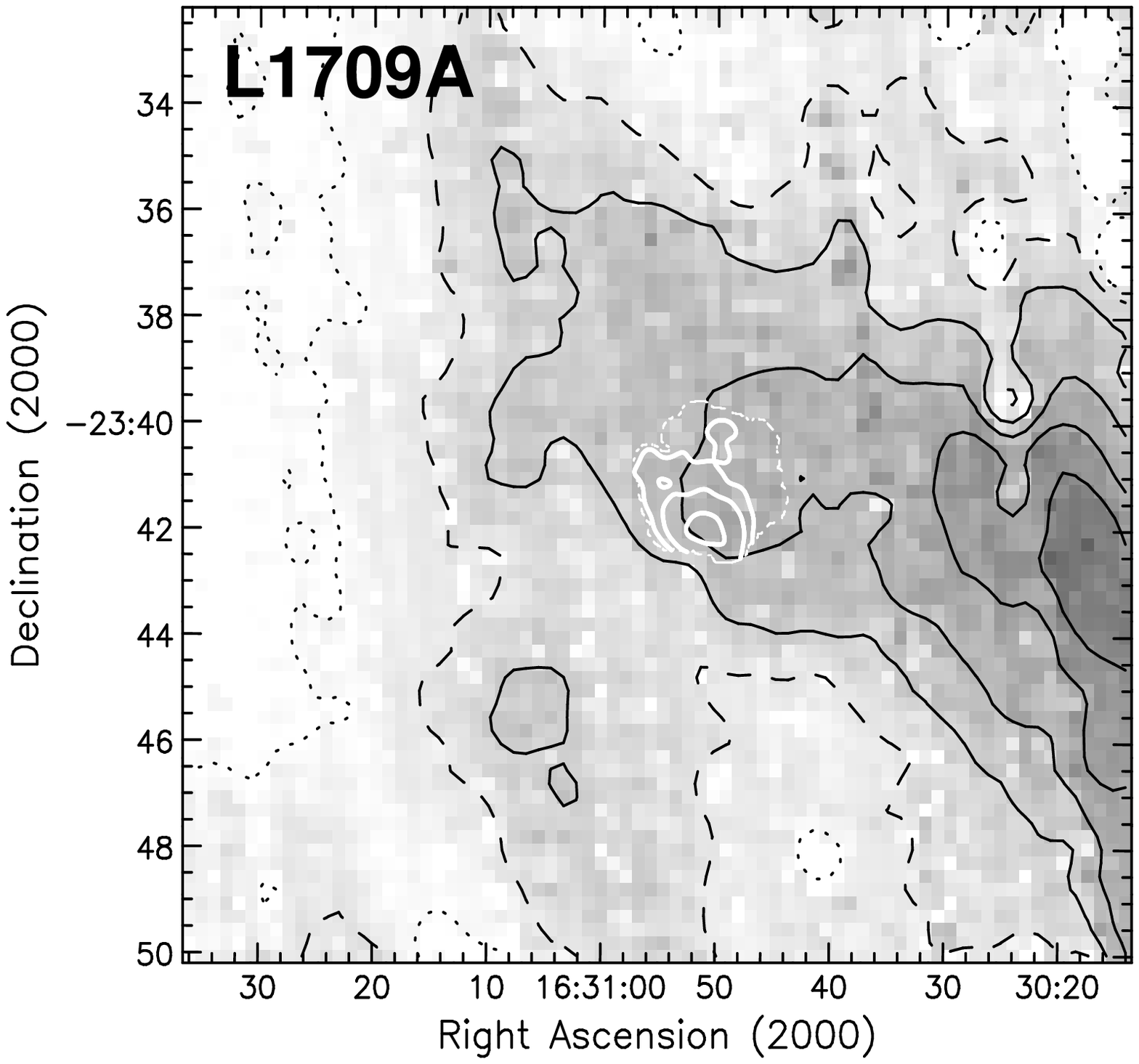} \\
		
	\end{tabular}
	\caption{Images of MIPS 160-\mum\ (greyscale with black contours) and SCUBA 850-\mum\ (white contours) continuum emission towards the pre-stellar cores L1498, L1521 (C, SMM, B, A, D, F, and E), L1517, L1512, L1696A and L1709A. The smoothed black 160-\mum\ contours start at the level of the on-cloud background (see Table \ref{tab:positions}) and proceed in plus (solid lines) and minus (dashed lines) 2-$\sigma$ increments. The lowest contour, the single dotted contour, marks the off-cloud background (see Table \ref{tab:positions}). The white 850-\mum\ contours are at 45, 75, and 95 per cent of peak intensity. The mapped SCUBA area is bounded by a dashed white line. Only this bound is shown for SCUBA non-detections.}
	\label{fig:cores1}
\end{figure*}

\begin{figure*}
	\begin{tabular}{ccc}
	\vspace{2mm}
		
		\includegraphics[height=0.28\textwidth]{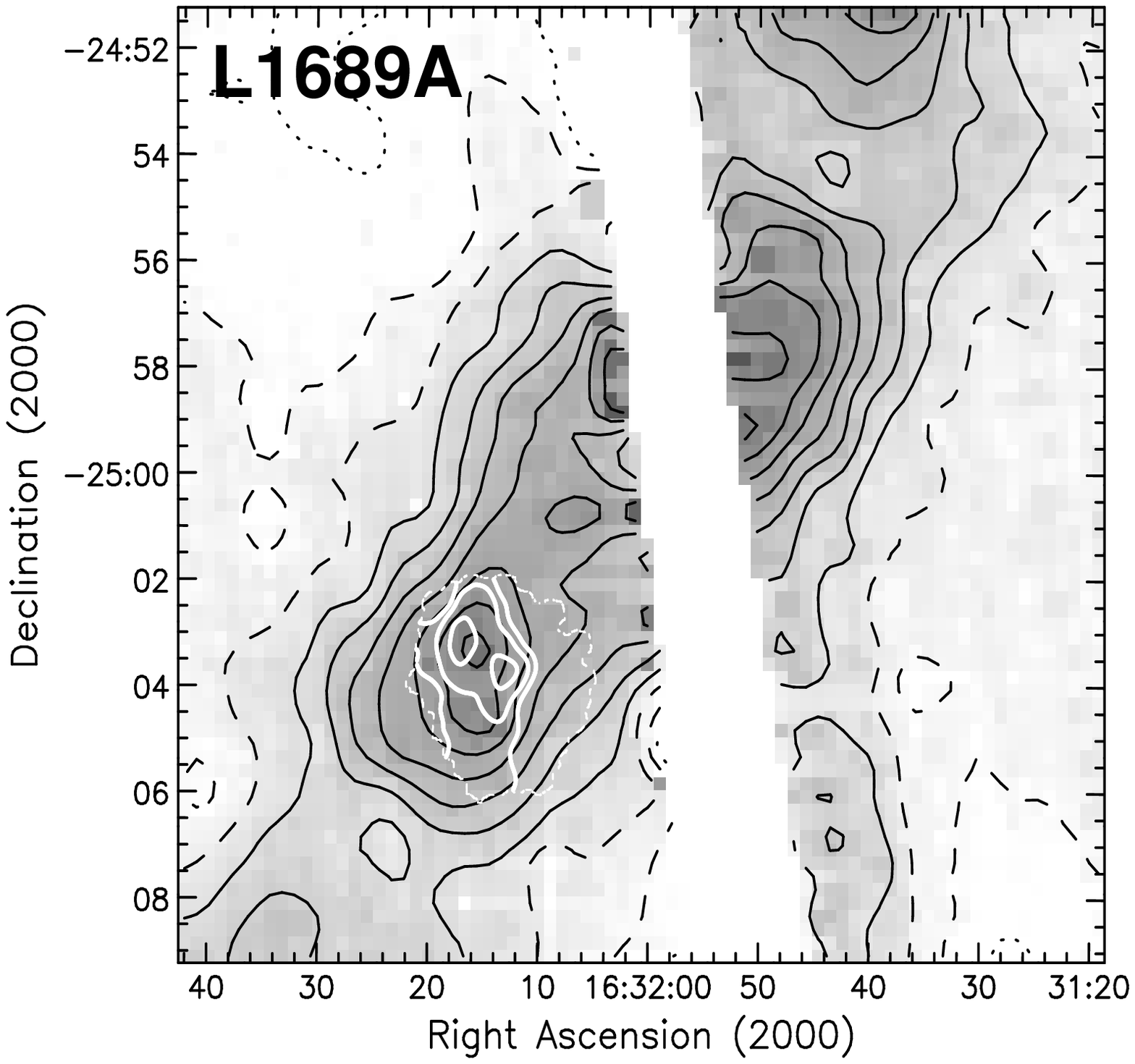} &
		\includegraphics[height=0.28\textwidth]{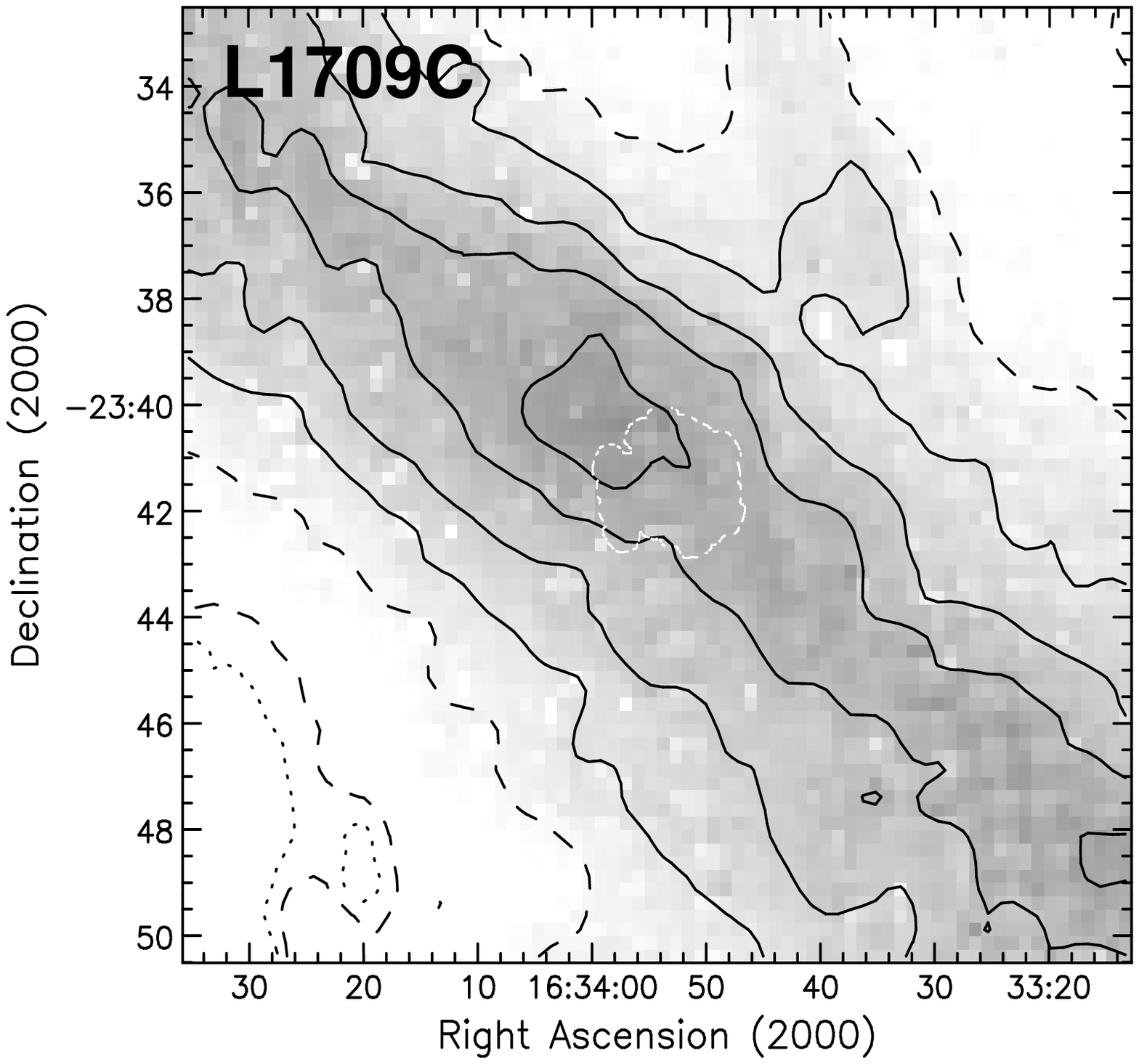} &		
		\includegraphics[height=0.28\textwidth]{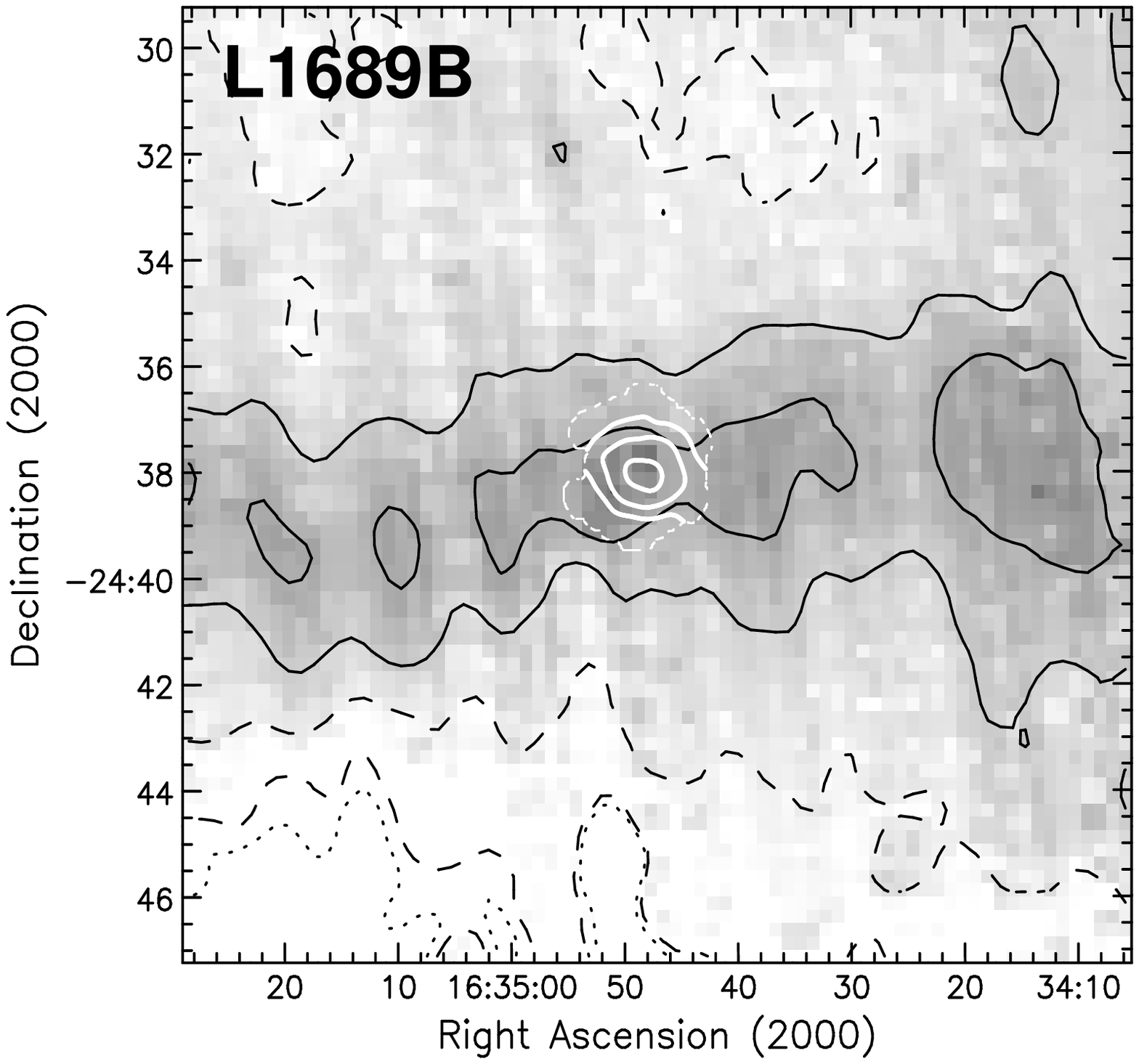} \\
		\multicolumn{3}{c}{
		\begin{tabular}{cc}
			\includegraphics[height=0.28\textwidth]{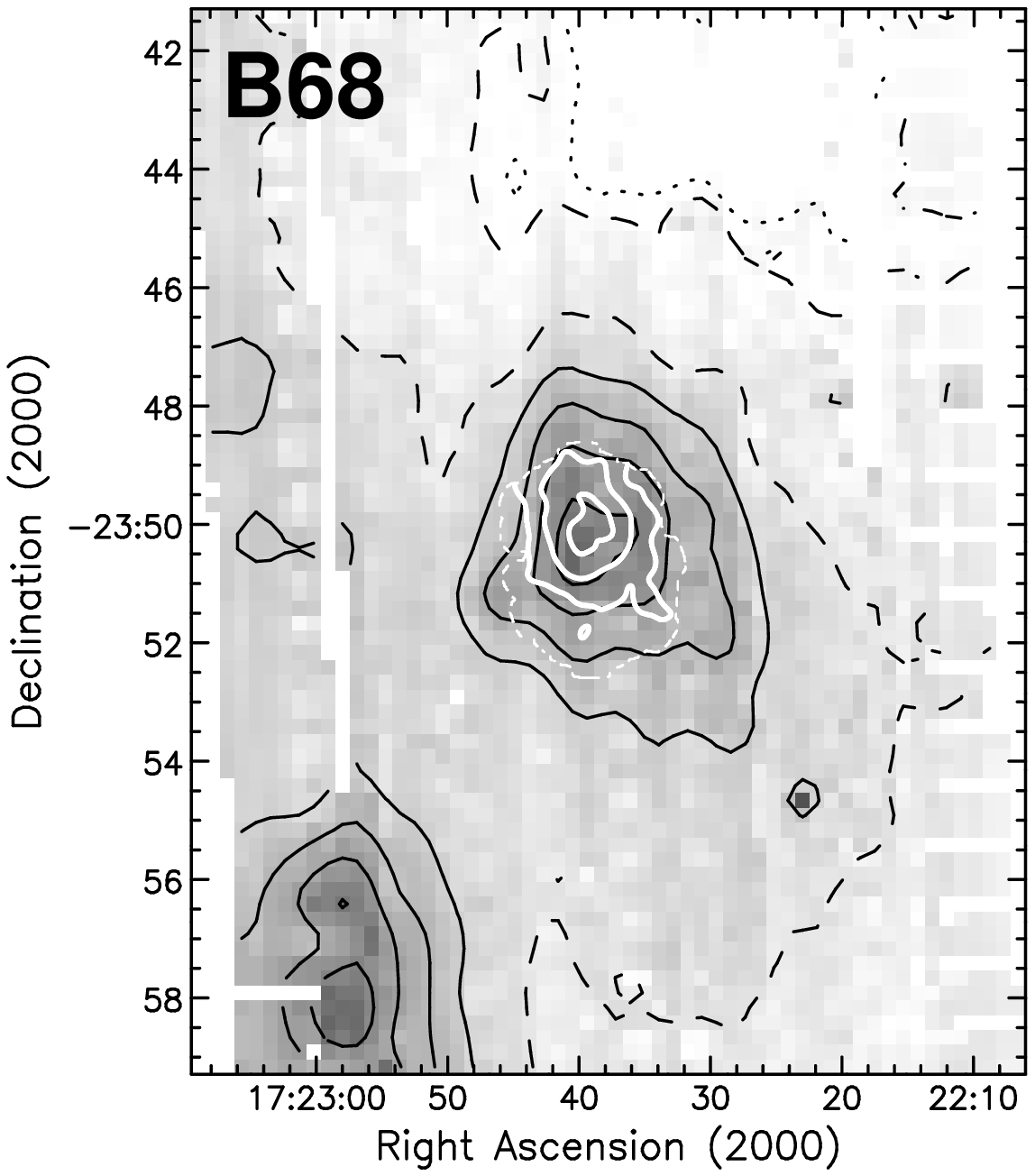} &
			\includegraphics[height=0.28\textwidth]{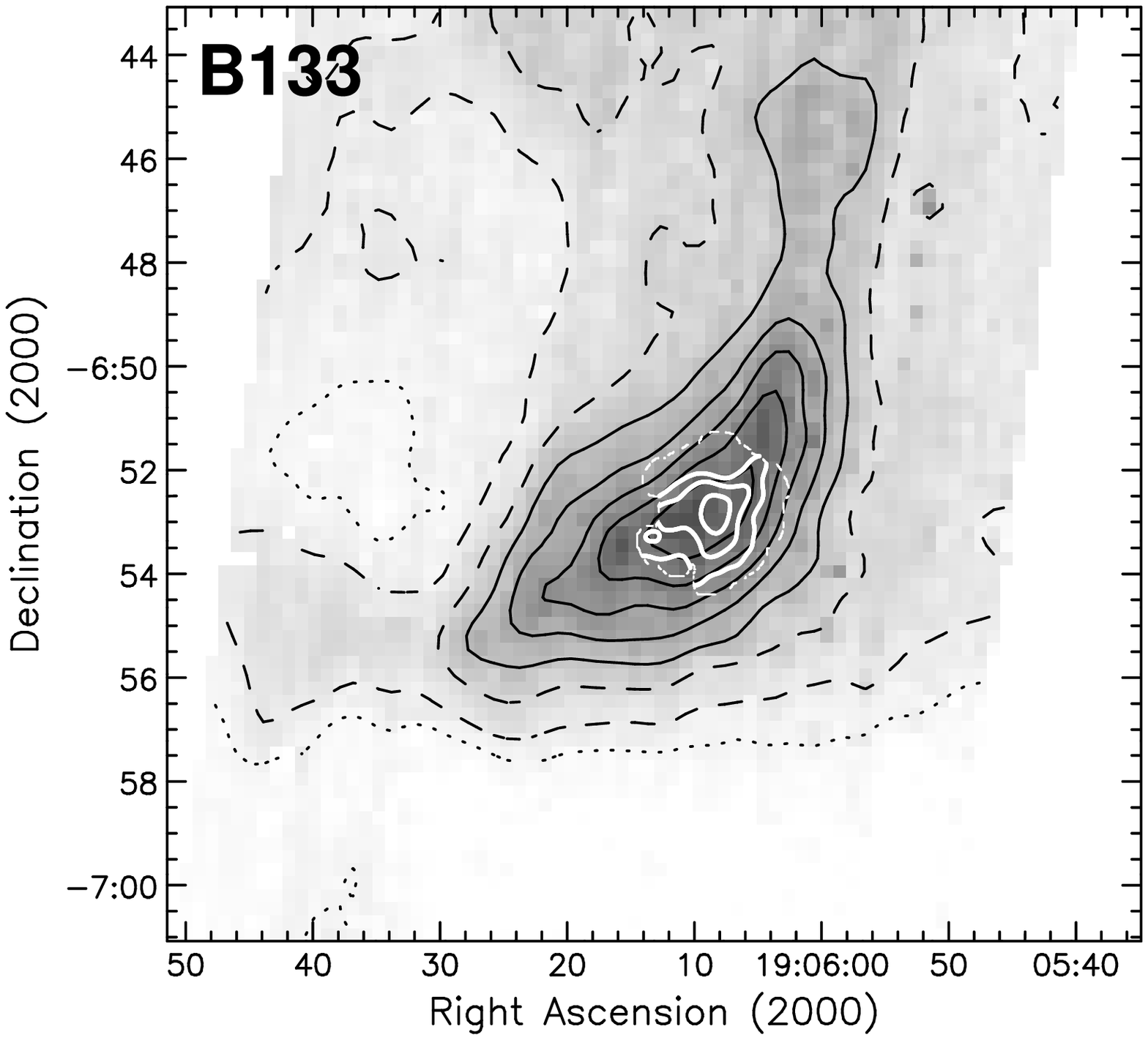}  \\
		\end{tabular}}
		
	\end{tabular}
	\caption{The pre-stellar cores L1689A, L1709C, L1689B, B68, and B133. Details as in Figure \ref{fig:cores1}.}
	\label{fig:cores2}
\end{figure*}

Figures \ref{fig:cores1} and \ref{fig:cores2} show the 18 cores that were downloaded from the \spitzer\ data archive. The greyscale shows the 160-\mum\ dust continuum and is scaled relative to the dynamic range of each field to avoid domination from strong sources. The dark contours plot the same continuum. The lowest solid contour is at the level as the on-cloud background (see Table \ref{tab:positions} and above), the other contours are then separated from it by increments of 2-$\sigma$ (solid above it, dashed below). The lowest contour is a dotted contour that shows the level of the off-cloud background. Superposed on each map are white contours that show the smoothed 850-\mum\ SCUBA continuum data from \citetalias{paper6} and \citet{2006kwc}, which cover a smaller area.

Almost all of our cores show strong 160-\mum\ structure that is correlated with the 850-\mum\ structure shown in the SCUBA data. Even those cores where SCUBA missed the emission peak show a surprising level of correlation, e.g. L1521C. The cores L1521A and L1709C were not detected with SCUBA, but the bounds of the observations are still shown in Figures \ref{fig:cores1} and \ref{fig:cores2}. The 160-\mum\ emission across those regions is relatively flat so it is possible that our non-detections were the result of there not being enough residual intensity to detect after the large scale emission had been chopped away.

An exception to the 850/160\mum\ pattern is the core L1696A (also known as Oph D). The small scale dumbbell-shaped SCUBA submillimetre emission closely matches that seen in absorption at 7\mum\ \citep{2000bacmann}, but it appears to be sat on a plateau of 160-\mum\ emission that has no discernible substructure. Detailed radiative transfer calculations suggest that the southern part of the core may be gravitationally bound \citep{2005steinacker}. 

\begin{figure*}
	\centering{\includegraphics[width=\textwidth]{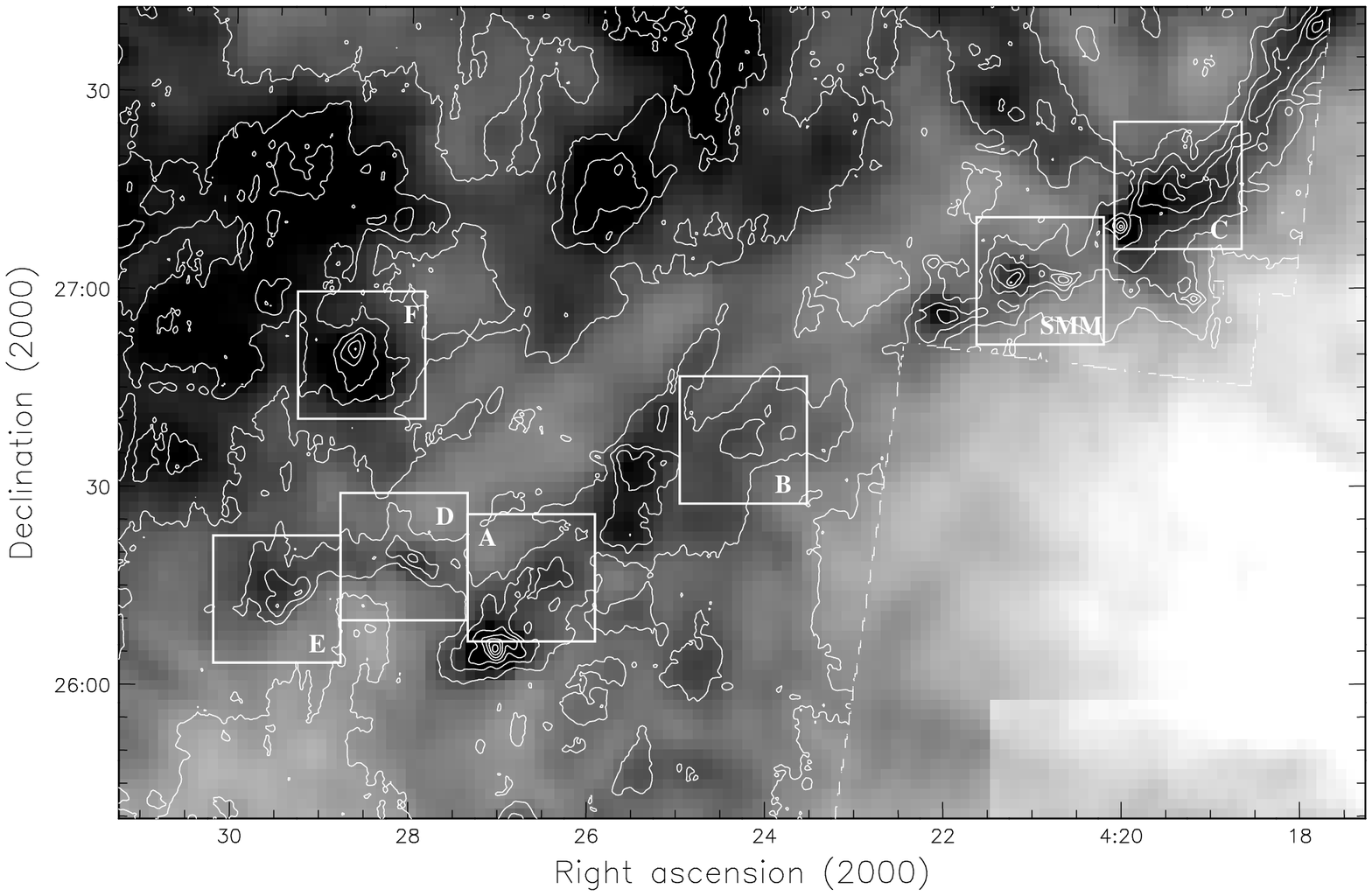}}
	\caption{The L1521 filament in Taurus. IRAS 100-\mum\ greyscale superposed with white MIPS 160-\mum\ contours. The contour heights start at 40 MJy/sr and proceed at intervals of 15 MJy/sr thereafter. The dashed white line shows the extent of the MIPS data. The seven boxes show the subregions shown in Figure \ref{fig:cores1} that are included within this area.}
	\label{fig:l1521}
\end{figure*}

L1521F, B68 and B133 all appear more isolated than the other cores, with strong central condensations whose closed contours extend almost to the edge of the maps. Half of the regions appear to be part of filamentary structures. Of the remaining images L1696A, L1498, and L1512 are on the edge of extended plateaux of emission while L1517 is a small cluster of star forming cores. L1517A and B are clearly seen at 160\mum, but the SCUBA non-detection of L1517D is now seen not to be a distinct core at 160\mum, so we discount it from our observations.

Seven of our sources are associated with the L1521 molecular filament that runs north-to-northwest from the west end of TMC-1 towards the L1495 region. A number of our fields touch each other so we have plotted their distribution in Figure \ref{fig:l1521}. The brightest of the L1521 cores, L1521F (also called MC 27), is separated from the main filament by half a degree and has been shown to be in an advanced evolutionary state very close to the time of protostar formation \citep{1999onishi,2005evo}. An embedded protostar has recently been reported in L1521F suggesting that it has reached the end of its pre-stellar phase \citep{2006bourke} and is a very young, low-luminosity Class 0 object somewhat similar to IRAM04191 also in Taurus \citep{1999amb}.

L1521E is a triple-lobed molecular core situated at the end of the L1521 filament. \citetalias{paper6} mapped the western lobe with SCUBA. This is the same lobe that has been identified as having a low chemical age based on comparative studies of molecular depletion \citep{2002hiy, 2004ts}.

The two SCUBA peaks seen in the L1521SMM field seem to be part of a chain of sources that run east-west across the map and are situated at the junction between the clouds 16a and 16b in the \citet{2002omkt} H$^{13}$CO$^{+}$ survey. The western-most SCUBA source is coincident with the position of a 2-\mum\ source identified as a brown dwarf, designated J04210795+2702204 \citep{2006gdmmm}. The eastern SCUBA source, which we denote as L1521SMM, is one of our pre-stellar cores from \citetalias{paper5}. The flux density from this source was measured in a 75-arcsec aperture to avoid confusion with the nearby object.

The map of L1689A shows a filament running diagonally across the map with a break in it caused by a gap in the surveyed area. This is labelled Filament 1 by \citet{2006nwa}. There are two pre-stellar cores from \citetalias{paper6} in this filament, but the L1689 SMM core lies under the coverage break. \citetalias{paper6} only partially mapped L1689A due to time constraints, but other SCUBA observations have shown that the core is double peaked \citep{2000serg,2006nwa}. We have added archival SCUBA data to our original map to produce the contours shown in Figure \ref{fig:cores2}. 

L1689B is an elongated core which has been shown to be sharp-edged in the north-south direction based on mm emission and mid-IR absorption maps \citep{paper2,2000bacmann}. The morphology seen in the \spitzer\ 160-\mum\ image supports the conclusion drawn from mm and mid-IR data.

B68 is a low-density circular core situated close to the Galactic Plane which has been identified as a critical Bonnor-Ebert sphere \citep{2001alves}. The bright source to the south-east is the starless core L55 SMM 1 \citep{2001vrc}.

\section{Spectral Energy Distributions}\label{seds}

\begin{table*}
	\caption{SCUBA positions and multi-band flux densities with 1-$\sigma$ errors used in the calculation of the dust SEDs. The 1-$\sigma$ errors include contributions from statistical noise and instrumental calibration errors (20 per cent for MIPS,  30 per cent for ISOPHOT, 25 per cent for SCUBA 450\mum, and 10 per cent for SCUBA 850\mum). It was found necessary to revise a couple of our SCUBA flux densities from \citetalias{paper6}. For example, slightly deeper 850-\mum\ polarimetry observations of L1498 showed that \citetalias{paper6} had under-estimated the flux density at 850\mum\ due to the orientation of the chop throw and the correction for this is included below \citep[see ][]{2006kwc}.  }
	\label{tab:seddata}
	\begin{center}
	\begin{tabular}{lcccccccccc}
		\hline
		Source	& Right Ascension & Declination	& 	\multicolumn{8}{c}{SED Flux Densities} \\
		& (2000)	& (2000) &	24-\mum &	70-\mum & 90-\mum & 160-\mum & 170-\mum & 200-\mum & 450-\mum & 850-\mum  \\
		& 		& 	 &	[mJy] &	[mJy] & [Jy] & [Jy] & [Jy] & [Jy] & [Jy] & [Jy]  \\

		\hline
	
		L1498	&	04$^{h}$ 10$^{m}$ 52.8$^{s}$   & $+$25\degr	10\arcmin 09\arcsec & $<$1.3 & $<$30 & $<$1.0 	& 7.0$\pm$1.4	& 10$\pm$3	& 15$\pm$5	& 13$\pm$3 	& 3.0$\pm$0.3  \\
		L1521SMM &	04$^{h}$ 21$^{m}$ 00.4$^{s}$   & $+$27\degr	02\arcmin 32\arcsec & --- & --- & --- & 0.86$\pm$0.20	& ---	& ---	& --- 	& 0.79$\pm$0.15  \\
		
		L1521D	&	04$^{h}$ 27$^{m}$ 47.0$^{s}$   & $+$26\degr	17\arcmin 38\arcsec & $<$3.3 & $<$61 &  --- 	& 3.4$\pm$0.7 	& --- 		& --- 		& 14$\pm$4 	& 3.1$\pm$0.3   \\
		L1521F	&	04$^{h}$ 28$^{m}$ 39.2$^{s}$   & $+$26\degr	51\arcmin 36\arcsec & 35$\pm$4 & 290$\pm$60 &  --- 	& 4.4$\pm$0.9 	& --- 		& --- 		& 12$\pm$4 	& 3.2$\pm$0.4   \\

		L1521E	&	04$^{h}$ 29$^{m}$ 13.6$^{s}$   & $+$26\degr	14\arcmin 05\arcsec & $<$2.6 & $<$67 &  --- 	& 0.66$\pm$0.23 & --- 		& --- 		& 4.0$\pm$1.4	& 1.4$\pm$0.2  \\
		L1517A  &	04$^{h}$ 55$^{m}$ 08.9$^{s}$   & $+$30\degr	33\arcmin 42\arcsec & $<$1.6 & 46$\pm$18 &  $<$1.2	& 5.8$\pm$1.2	& 9.6$\pm$2.9	& 14$\pm$5	& 4.8$\pm$1.3	& 0.73$\pm$0.14  \\
		L1517B	&	04$^{h}$ 55$^{m}$ 17.9$^{s}$   & $+$30\degr	37\arcmin 47\arcsec & $<$1.6 & $<$46 &  $<$0.78 	& 3.4$\pm$0.7 		& 5.9$\pm$1.8 	& 8.4$\pm$3.2 	& 12$\pm$3 	& 2.6$\pm$0.3  \\

		L1512	&	05$^{h}$ 04$^{m}$ 08.4$^{s}$   & $+$32\degr	43\arcmin 26\arcsec & $<$1.4 & 41$\pm$12 &  $<$1.4 	& 9.9$\pm$2.0 	& 12$\pm$4 	& 16$\pm$5 	& 8.0$\pm$2.3 	& 1.4$\pm$0.2  \\

		L1696A	&	16$^{h}$ 28$^{m}$ 28.9$^{s}$   & $-$24\degr	19\arcmin 09\arcsec & $<$13 & $<$590 &  $<$2.1 	& 38$\pm$8 	& 39$\pm$12 	& 48$\pm$14	& --- 		& 6.3$\pm$0.6  \\
		L1709A	&	16$^{h}$ 30$^{m}$ 51.9$^{s}$   & $-$23\degr	41\arcmin 52\arcsec & $<$4.4 & $<$250 &  $<$1.6 	& 8.9$\pm$1.8 	& 10$\pm$4 	& 18$\pm$5 	& --- 		& 1.5$\pm$0.2  \\
		L1689A	&	16$^{h}$ 32$^{m}$ 13.2$^{s}$   & $-$25\degr	03\arcmin 45\arcsec & $<$4.0 & 13000$\pm$3000 &  35$\pm$10 & 130$\pm$30 	& 130$\pm$40 	& 120$\pm$40	& 19$\pm$6 	& 2.4$\pm$0.3  \\
		L1689B	&	16$^{h}$ 34$^{m}$ 48.2$^{s}$   & $-$24\degr	38\arcmin 04\arcsec & $<$15 & $<$330 &  --- 	& 35$\pm$7	& 29$\pm$9 	& 29$\pm$9 	& 18$\pm$6 	& 3.1$\pm$0.4  \\
		
		B68	&	17$^{h}$ 22$^{m}$ 39.2$^{s}$   & $-$23\degr	50\arcmin 01\arcsec & $<$2.3 & $<$75 &  $<$2.8 	& 12$\pm$2 	& 12$\pm$4 	& 16$\pm$5 	& $<$8.5 	& 1.3$\pm$0.2  \\
		B133	&	19$^{h}$ 06$^{m}$ 08.4$^{s}$   & $-$06\degr	52\arcmin 22\arcsec & $<$2.0 & 110$\pm$30 &  $<$4.5 	& 19$\pm$4 	& 23$\pm$7 	& 35$\pm$10 	& 16$\pm$5 	& 1.7$\pm$0.2  \\
		\hline
	\end{tabular}
	\end{center}
\end{table*}

We can add 160-\mum\ flux density measurements to our spectral energy distributions from \citetalias{paper5}. However we are limited by the extent of our SCUBA maps and cannot use the centroid flux densities presented in Table \ref{tab:positions}, as they are not always coincident with the SCUBA positions. It is therefore necessary to measure flux densities in a common aperture centred on a common position for all wavelengths. 

Table \ref{tab:seddata} lists the 14 sources common to \citetalias{paper5}, \citetalias{paper6}, and this paper, the positions of the SCUBA cores and their multiband flux densities measured in a 150 arcsec aperture centred on that position. Columns 6, 8 and 9 list the ISOPHOT 70, 170, and 200-\mum\ flux densities from \citetalias{paper5}, table 4. Column 7 lists the \spitzer\ 160-\mum\ flux density measured at the SCUBA position (note this is a different position and flux density than listed in Table \ref{tab:positions}). Columns 10 and 11 list the SCUBA 450 and 850-\mum\ flux densities from \citetalias{paper6}, tables 1 and 2. Four sources, L1521A, B \& C, and L1709C were either undetected or missed by SCUBA. 

In order to enhance our short wavelength coverage, background subtracted flux density measurements were made at the SED positions using 24- and 70-\mum\ \spitzer\ data. These were downloaded from the \spitzer\ archive. C2D legacy data were used for the Ophiuchus regions (R.A.=16$^{h}$). Of the 14 pre-stellar cores with \spitzer\ data only 5 were detected at 70\mum\ and only L1521F was detected at 24\mum. Columns 4 and 5 of Table \ref{tab:seddata} list the 24- and 70-\mum\ flux densities and 1-$\sigma$ uncertainties or, where appropriate, 3-$\sigma$ upper limits. 

At 24-\mum\ several of the Ophiuchus cores were silhouetted against brighter background emission. The 3-$\sigma$ upper-limits derived from the 24-\mum\ images varied according to the region and the scan speed. A typical 24-\mum\ 3-$\sigma$ upper-limit for an isolated Taurus core observed using a medium-scan speed was $\sim$1.6mJy (e.g. L1517B), whereas the 24-\mum\ 3-$\sigma$ upper-limit for some Ophiuchus cores observed using a fast-scan speed was an order of magnitude higher (e.g. L1689B). 

\begin{figure*}
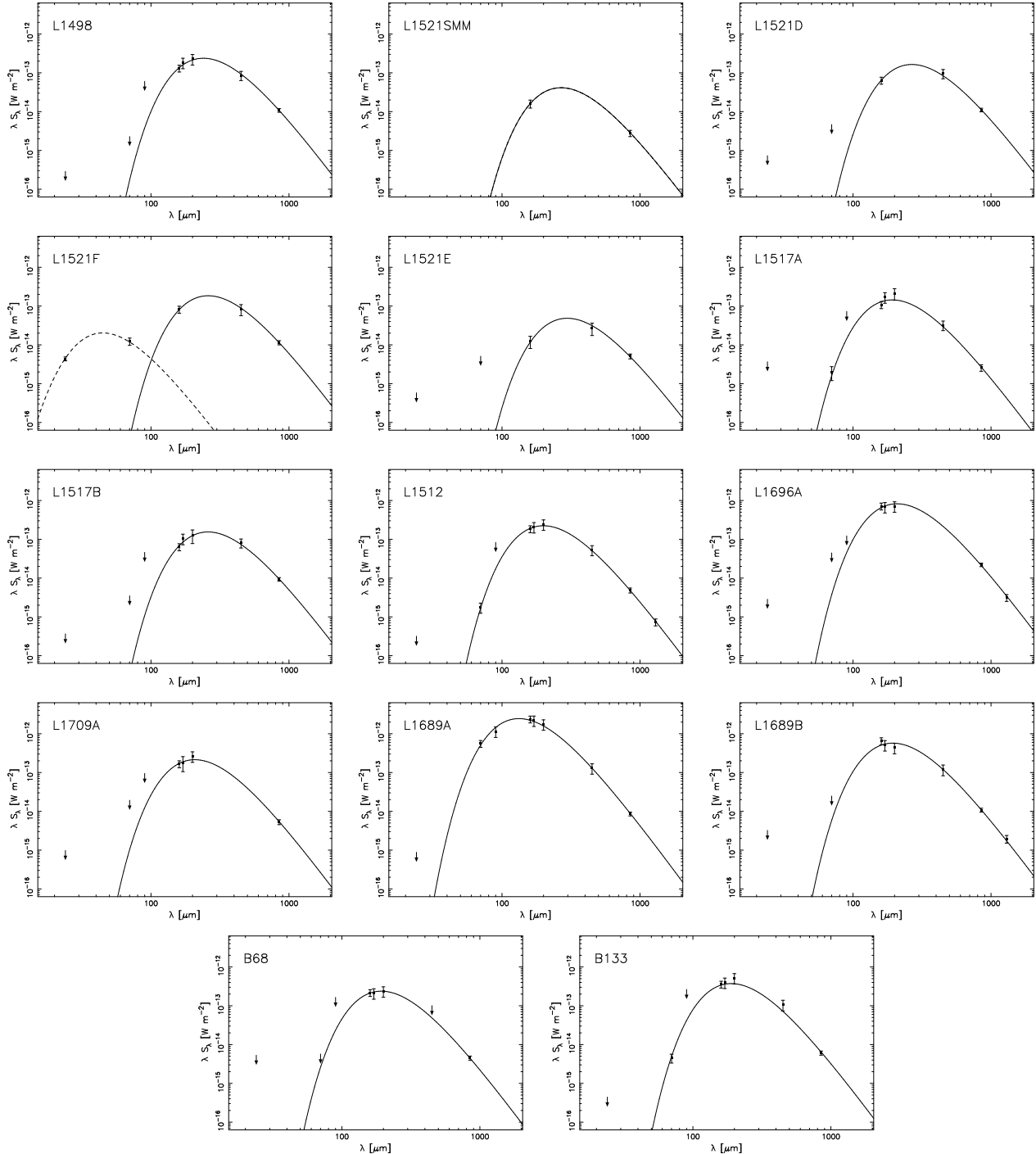

	\begin{tabular}{ccc}
		\includegraphics[width=0.2\textwidth,angle=-90]{L1498_lambdaFlambda.ps} &
		\includegraphics[width=0.2\textwidth,angle=-90]{L1521SMM_lambdaFlambda.ps} &
		\includegraphics[width=0.2\textwidth,angle=-90]{L1521D_lambdaFlambda.ps} \\
		
		\includegraphics[width=0.2\textwidth,angle=-90]{L1521F_lambdaFlambda.ps} &
		\includegraphics[width=0.2\textwidth,angle=-90]{L1521E_lambdaFlambda.ps} &
		\includegraphics[width=0.2\textwidth,angle=-90]{L1517A_lambdaFlambda.ps} \\
		
		\includegraphics[width=0.2\textwidth,angle=-90]{L1517B_lambdaFlambda.ps} &
		\includegraphics[width=0.2\textwidth,angle=-90]{L1512_lambdaFlambda.ps} &
		\includegraphics[width=0.2\textwidth,angle=-90]{L1696A_lambdaFlambda.ps}\\
		
		\includegraphics[width=0.2\textwidth,angle=-90]{L1709A_lambdaFlambda.ps} &
		\includegraphics[width=0.2\textwidth,angle=-90]{L1689A_lambdaFlambda.ps} &
		\includegraphics[width=0.2\textwidth,angle=-90]{L1689B_lambdaFlambda.ps} \\

		\multicolumn{3}{c}{
			\begin{tabular}{cc}	
				\includegraphics[width=0.2\textwidth,angle=-90]{B68_lambdaFlambda.ps} & 	
				\includegraphics[width=0.2\textwidth,angle=-90]{B133_lambdaFlambda.ps}  \\
			\end{tabular}
		}

	\end{tabular}
	\caption{Spectral Energy Distributions for the cores listed in Table \ref{tab:seddata}. Each plot, where available, shows the flux densities and 1-$\sigma$ calibration uncertainties or 3-$\sigma$ upper-limits at 1.3mm (IRAM), 850 and 450-\mum\ (SCUBA), 200, 170, and 90-\mum\ (ISOPHOT), and 160, 70 and 24-\mum\ (MIPS). The solid curves are greybody fits to the data. A dashed curve indicates the presence of a second hotter dust component in the greybody fit. This is only seen in one core, L1521F, now believed to be protostellar, where the warmer component has $T=63\pm5$K.}
\label{fig:seds1}
\end{figure*}

The data from Table \ref{tab:seddata} and selected data from \citetalias{paper3} are plotted on log-log plots as a function of wavelength in Figure \ref{fig:seds1}.  Fitted against the data from Table \ref{tab:seddata} and plotted in Figure \ref{fig:seds1} are a series of modified blackbody, or greybody curves. The monochromatic flux density $S_{\nu}$ of a greybody, at frequency $\nu$, radiated into solid angle $\Omega$ is given by
\begin{equation}
S_{\nu} = \Omega f B_{\nu,T} [ 1- e^{-(\frac{\nu}{\nu_{c}})^{\beta}} ],
\end{equation}
where $B_{\nu,T}$ is the Plank function, $\nu_{c}$ is the frequency at which the optical depth is unity, $\Omega$ is the solid angle of the aperture, $f$ is the filling factor of the source within the aperture and $\beta$ is the dust emissivity index. A greybody was fitted through each set of data points by using $\chi^{2}$-minimisation. It was assumed that $\beta$ = 2 and that $\nu/\nu_{c}=1$ at 50 \mum, after \citet{paper5}. The parameters $f$ and $T$ were then mapped on a 1000$\times$1000 grid of $\chi^{2}$.  Typical 1-$\sigma$ deviations from the best fit were found to be between 0.3 and 0.5K.  

\begin{table}
	\caption{SED fitting results. Columns 2, 3 and 4 list the temperature, peak wavelength, and luminosity (between 30\mum\ and 3mm) of the best fit SED. Column 5 lists the reduced $\chi^{2}$ parameter for the SED fit quoted to 1 s.f. It was assumed that $\beta$=2. Column 6 lists the equivalent SED temperatures from \citetalias{paper5}. L1521SMM has too few data points for a sensible determination of the $\chi^{2}$ parameter, but its data are consistent with the SED of a 9K greybody. L1521F has a second hotter dust component as shown in Figure \ref{fig:seds1}, columns 3-5 list parameters for the colder of the two components while the $\chi^{2}$ parameter listed in Column 5 is for the combined fit. }
	\label{tab:sedfit}
	\begin{center}
	\begin{tabular}{lccccc}
		\hline
		Source		& $T$ 	& $\lambda_{peak}$ & $L_{IR-MM}$ &	$\chi^{2}_{\nu}$ & T$_{ISO}$  \\
				& [K]	& [\mum] & [L$_{\odot}$] & & [K]  \\ 
		\hline
		L1498		& 10.1$\pm$0.3 	& 290	& 0.15	& 0.1 	& 10 	\\
		L1521SMM	&  9.0$\pm$1.0	& 320	& 0.03  & ---	& ---	\\
		L1521D		&  9.0$\pm$0.3 	& 320	& 0.11	& 0.2 	& --- 	\\
		L1521F 		&  9.3$\pm$0.3 	& 310 	& 0.12	& 0.1 	& ---	\\
		L1521E		&  8.1$\pm$0.4 	& 360	& 0.031	& 0.3 	& --- 	\\
		L1517A		& 12.6$\pm$0.4  & 240	& 0.093	& 0.6	& --- 	\\
		L1517B		&  9.3$\pm$0.3 	& 310	& 0.10	& 0.2 	& 10 	\\
		L1512		& 12.1$\pm$0.3	& 240	& 0.14	& 0.04 	& --- 	\\
		L1696A		& 11.6$\pm$0.4 	& 250	& 0.46	& 0.3 	& 10 	\\
		L1709A		& 11.6$\pm$0.5 	& 250	& 0.12	& 0.2 	& ---	\\
		L1689A		& 18.5$\pm$0.5 	& 160	& 1.4	& 0.3 	& 19 	\\
		L1689B		& 12.5$\pm$0.5	& 240	& 0.32	& 0.6 	& 11	\\
		B68		& 12.5$\pm$0.5 	& 240	& 0.13	& 0.04 	& 10	\\
		B133		& 12.8$\pm$0.3 	& 230	& 0.49	& 0.6 	& 13 	\\
		\hline
	\end{tabular}
	\end{center}
\end{table}

Our best-fit parameters are listed in Table \ref{tab:sedfit} along with the original ISOPHOT fit temperature from \citetalias{paper5}. The agreement between the two temperatures is seen to be good. The slight difference between the two is due to the addition of the 160-\mum\ point which gives greater weight to the fit around the peak. There are no detected systematic differences between the original and updated temperatures. Column 3 of Table \ref{tab:sedfit} lists the wavelength of peak emission, $\lambda_{peak}$, for each best fit SED. Column 4 lists the infra-red to mm luminosity, $L_{IR-MM}$ derived by integrating the best-fit SEDs between 30\mum\ and 3mm.  

We have both ISOPHOT 170\mum\ and MIPS 160\mum\ flux densities for ten cores. All except one of the 160\mum\ flux densities are either comparable with or slightly less than the 170\mum\ flux densities, in broad agreement with what would be expected for a 10--13K SED. The agreement in flux densities for the Ophiuchus cores is particularly interesting as their peak intensities from Table \ref{tab:positions} would indicate they are heavily into the MIPS saturated regime. 

The SED temperatures for L1521D and L1521E are approximately 9K, validating our assumption in \citet{paper6} of thermal equilibrium between the dust temperature and the thermal rotational line temperatures measured by \citet{1997codella}. The temperatures for the three L1521 cores are slightly lower than for the other cores. L1521F shows a short wavelength excess above the single cold SED fitted to the rest of the cores. A second hotter greybody can account for this excess as shown in figure \ref{fig:seds1}. This second component has a temperature of 63$\pm$5K and a luminosity of $\sim$ 0.01 L$_{\odot}$ making it one tenth as luminous as the extended core. This is consistent with the  properties of the protostellar object detected in L1521F by \citet{2006bourke}.

The mean temperature of these prestellar cores is 12K and is significantly lower than the mean temperature (23K) of equivalent isolated cores with an embedded protostar \citep{1997launhardt}.

\section{Discussion}\label{discuss}

\subsection{Core Classification}

Recently there have been reports of very low luminosity ($L\le$ 0.1 L$_{\odot}$) protostellar objects embedded within molecular cores that were previously believed to be pre-stellar \citep{2005crapsi,2006bourke}. These recently discovered protostars are similar to some previously known Class 0 protostars such as VLA 1623 \citep{1993awb}, HH24MMS \citep{1995wckab,1996bwa}, and IRAM 04191 \citep{2006dunham}. The core L1521F is now believed to harbour a young protostellar object, based on its detection by \spitzer\ at 24\mum\ \citep{2006bourke}. A detection at this wavelength seems to be key in identifying embedded protostars in dense cores that were thought to be pre-stellar, due to the great sensitivity of \spitzer\ in the mid-infrared. 

We have confirmed the L1521F detection above and shown that this requires there to be hotter dust within L1521F -- a signature of an internal heating source. We can determine from our combined \spitzer\ and SCUBA data whether any other pre-stellar cores contain warm dust, and therefore may also contain young protostars.

Of the 29 pre-stellar cores detected by SCUBA at 850-\mum\ in \citetalias{paper6}, 23 had coincident MIPS obervations present in the \spitzer\ data archive, and of those 14 had usable 160-\mum\ data (as represented by the SEDs in Section \ref{seds}). One of the remaining cores, L1524, is actually protostellar and had been misclassified. The remaining eight cores are L1686, L43, L234E, L63, L1148, L1155C, H and D. 

\citet{2005kauffmann} identified a protostellar candidate in L1148. However, their source, L1148-IRS, is not coincident with our SCUBA detection and is a separate neighbouring object. We suggest the name L1148-SMM for the prestellar core to differentiate if from the protostellar core L1148-IRS. The separation between L1148-IRS and L1148-MMS is 200 arcsec, which equates to $\sim$0.3 pc given a distance to L1148 of 325 pc \citep{1992straizys}. Archival and C2D legacy 24-\mum\ maps of the remaining 8 cores were checked for the presence of embedded protostars, but no obvious candidates were found. 

\citet{2005swbg} defined three criteria for differentiating young protostellar cores from pre-stellar cores based on their SEDs ($T>15$K, $\lambda_{peak}<170$ \mum) and submillimetre morphology (approximately circular within 4000AU). One of our cores, L1689A, satisfies both conditions. This is also the core whose SED peaks furthest from the radio and which is most luminous. L1689A has two peaks in the submillimetre and is more condensed than the other cores. Based on the criteria listed by \citet{2005swbg}, L1689A could have been considered a potential protostellar candidate, but it was not detected by \spitzer\ at 24\mum, and so does not appear to contain any warm dust. Likewise, no other cores in our sample other than L1521F were detected at 24\mum within the cores' full-width at half-maximum (FWHM, see Table \ref{tab:coreprop}) at 160\mum. 

Therefore, out of the 22 cores common to our \spitzer\ and SCUBA samples that were previously believed to be pre-stellar, only one contains a protostar. This appears to indicate that very low luminosity protostellar objects embedded within pre-stellar cores are rare contrary to some early suggestions \citep[e.g.][]{2004young}.

We used the method of statistical lifetimes \citep{1986beichman,1999lee} to estimate the lifetime of a prestellar core. In short, this method uses the relative abundance of sources to determine their relative statistical lifetimes in a general source population (the rarer an object the shorter its lifetime). One caveat to this method is that it requires a known lifetime for the entire system or of one part of it in order to calibrate the rest of the inferred lifetimes. For starless cores the reference lifetime of $\sim6\times10^{5}$ years is inferred from the lifetime of the Class I protostellar phase \citep[see][and references therein]{paper6}. 

From the starless core lifetime we estimated that the statistical lifetime for a prestellar core detectible in the submillimetre was $\sim3.0\times10^{5}$ years \citep{paper6}. This was based on a sample of 27 cores detected out of 50. A misclassification of only 1 out of 22 cores, as shown above, would shorten this estimate by $\sim$5 per cent, which is well within the errors associated with this lifetime \citep[for a discussion see][]{2006wtppv}. Hence we see no reason to adjust the pre-stellar core lifetime estimate based on the new \spitzer\ data.

\subsection{Mean Core Properties}

\begin{table}
	\caption{Table of mean core statistics derived from the MIPS and SCUBA data. Column 2 lists the mean value derived from the \spitzer\ 160-\mum\ observations for the thirteen prestellar cores discovered in this paper. Column 2 lists the equivalent mean parameter derived from the SCUBA 850-\mum\ observations in Table 4 of \citet{paper6} (except L1521F and L1582 whose greater distance skews the mean). The values in parenthesis are for the 7 cores common to each sample. A ``bg'' subscript denotes a quantity that has had the background intensity subtracted. }
	\label{tab:coreprop}
	\begin{center}
	\begin{tabular}{lccl}
		\hline
			 			& \spitzer\ 160\mum\ & SCUBA 850\mum\ & \\
		\hline
		Sample size			& 13 (7)	& 10 (7)	& \\
		$I_{\nu}$ 			& 230 (170)	& ---		& MJy/sr \\
		$I_{\nu,\mathrm{bg}}$ 		& 86 (57) 	& 34 (27)	& MJy/sr \\
		$S_{\nu,\mathrm{bg}}$ 		& 22 (13) 	& 2.7 (2.2)	& Jy \\
		$M_{\mathrm{150 arcsec},bg}$ 	& 1.4 (1.5) 	& 1.6 (1.3)	& M$_{\odot}$ \\
		$N($H$_{2})_{\mathrm{peak}}$	& 7.8 (5.4) 	& --- 		& $10^{22}$ cm$^{-2}$ \\
		$N($H$_{2})_{\mathrm{peak,bg}}$	& 1.7 (1.4) 	& 4.5 (3.1)	& $10^{22}$ cm$^{-2}$ \\
		FWHMa 				& 0.125 (0.140) & 0.051 (0.058)	& pc	\\
		FWHMb				& 0.084 (0.086) & 0.028	(0.033)	& pc	\\
		HWHM$_{mean}$			& 0.051 (0.054) & 0.016 (0.022) & pc \\
		Aspect Ratio			& 0.67 (0.69) 	& 0.59 (0.61)	&	\\
		$M_{\mathrm{FWHM},bg}$ 		& 1.7 (1.7) 	& 0.5 (0.4)	& M$_{\odot}$ \\
		$r_{BE}$			& ---		& $\sim$0.150	& pc \\
		$r_{\mathrm{cc}}$		& $\sim$0.120	& ---		& pc \\
		\hline
	\end{tabular}
	\end{center}
\end{table}

Table \ref{tab:coreprop} shows the mean core properties derived from our SCUBA observations \citepalias[table 4]{paper6} and our \spitzer\ observations (this paper). The first line lists the sample size with the numbers in parenthesis referencing the 7 cores that are common to both samples (L1521F has been excluded from both). Lines 2--4 list the measured peak intensities and flux densities in a 150 arcsec aperture. A ``bg'' subscript denotes a background subtracted measurement or a quantity derived from that measurement. SCUBA observations are automatically counted as background subtracted because it uses a chopping secondary mirror to reject sky emission. 

Line 5 lists the mean of the core masses derived from the 150 arcsec aperture fluxes at the two wavelengths. When constructing an SED we assume that the same mass of gas is radiating at all wavelengths. If that assumption holds then we should expect to derive identical mass estimates from different points on the SED. That the two masses on this line are virtually identical is apparent proof of the aforementioned assumption. 

Lines 6 and 7 of Table \ref{tab:coreprop} list the peak H$_{2}$ column density before and after background subtraction. If the core is centrally condensed then we could expect the smaller 850-\mum\ beam to preferentially sample a higher mean column density than the more extended 160-\mum\ beam, which it does.

Lines 8 and 9 list the mean deconvolved major and minor full-width at half-maximum (FWHM) of the cores. For 160\mum\ this is the half-maximum between the peak and the on-cloud background. Line 10 lists HWHM$_{mean}$, the geometric mean of half of the FWHM ellipse and line 11 lists its aspect ratio. Line 12 lists $M_{\mathrm{FWHM},bg}$ the mean estimated mass enclosed by the FWHM ellipse. The $M_{\mathrm{150 arcsec},bg}$ and $M_{\mathrm{FWHM},bg}$ are nearly identical at 160\mum\ because the mean core diameter translates to an angular size of $\sim$150 arcsec at the typical distance of these cores (140 parsec). 

\citet{2006wtppv} defined prestellar cores as ``that subset of starless cores which are gravitationally bound and hence are expected to participate in the star formation process''. For the prestellar cores observed with SCUBA $M_{vir} \sim M_{obs}$ \citepalias{paper6} and can therefore be considered bound. Between the two wavelengths shown in Table \ref{tab:coreprop} the observed radius and mass increase by a factor of three. For a virialised cloud $M \propto \Delta v^{2} R$ where $\Delta v$ is the turbulent linewidth of the gas. Therefore, for a constant linewidth $M$ is linear with respect to $R$ \citep[c.f.][]{larson81}. This is the same as the linear scaling seen between the cores as observed with SCUBA and as observed with \spitzer\ so it is reasonable to assume that the extra material seen by \spitzer\ is also in bound. 

When comparing the FWHM size and mass of the cores we note that the mean SCUBA sizes and masses are in agreement with other SCUBA size/mass studies of nearby star formation \citep[e.g.][]{1998man} whilst the mean of the Spitzer sizes and masses are close to submillimetre studies of condensations in the Orion region \citep[e.g.][]{2001motte,2006nw}.

With these observations and our SCUBA observations we are able to probe a spatial regime down to 2000 AU (the FWHM of the SCUBA 850\mum\ at 140 parsecs). For the 13 confirmed prestellar cores we predominantly see a one to one mapping of closed SCUBA contours onto closed \spitzer\ contours suggesting that we are detecting the same single coherent unfragmented object in both sets of observations. Most stars form in multiple systems \citep[see][and references there in]{2006goodwin} with the implication that at some scale a single core must fragment into multiple protostellar fragments. That we see no such fragmentation over the scales that we are sensitive to supports the finding of \cite{2004goodwin} that fragmentation must occur below the 2000AU scale and/or at a later core evolutionary stage (such as the Class 0 stage).

Fitting of Bonner-Ebert (BE) spheres to SCUBA maps gives an implied pressure bounded outer radius of $r_{BE} = 3\times10^{4}$ AU (0.150 parcecs; line 13 of Table \ref{tab:coreprop}) \citep{evans-cores2,paper6}. This is approximately equal to the major FWHM of the \spitzer\ cores and the radius of dense cores mapped in C$^{18}$0 \citep{1991myers,1998onishi}. The mean HWHM of cores in the \spitzer\ observations is approximately equal to the size of prestellar cores measured from N$_{2}$H+ \citep[0.054 parsecs;][]{2002n2hp} and H$^{13}$CO+ \citep[0.048 parsec;][]{2002omkt}. \citet{2006young} used the SCUBA scan mapping mode to map larger areas over a selection of prestellar and protostellar cores. Their maps showed that the area of emission detected from cores with SCUBA was generally smaller than 2 arcmin across (the width of the SCUBA field of view).

From the above radii we can define a ``characteristic'' radius for the full extent of a prestellar core. This is $\sim$0.15 parsec and is the BE pressure bounded radius and the radius as measured from CO observations. At the distance to these sources this radius is equivalent to $\sim 3.5$ arcmin. Given the complexity of the lowest on-cloud contour in Figures \ref{fig:cores1} and \ref{fig:cores2} it is nearly impossible to measure a full extent radius. However, by inspection, the regions typically have a closed contour radius, $r_{\mathrm{cc}}$, of $\sim$3 arcmin (0.12 parsecs; line 14 of Table \ref{tab:coreprop}), comparable to the 3.5 arcmin BE radius. Then, within this bounding radius there is a centrally condensed region of radius $\sim 0.05$ parsec as delineated by observations of high density gas tracers and the FWHM of far-IR dust emission. 

\section{Conclusion}

In this paper we have presented \spitzer\ archive data of a sample of 18 molecular cloud cores that were believed to be pre-stellar in nature. The sample was selected from the 52 cores that we had previously observed with SCUBA. We have produced new images of the cores at 160\mum\ and we have measured their flux densities at the other MIPS wavelengths of 24 and 70\mum\ to produce full SEDs for the cores, when combined with our previous data. Comparison of background intensities and measured flux densities showed consistency between the absolute and relative MIPS 170\mum and ISOPHOT 160\mum calibration. Four of the cores were non-detections in our SCUBA sample. We refined our previous fits to the remaining 14 core SEDs and produced temperature estimates for the cores. Most lie in the range 10--13K. Only one core was detected at 24\mum\ and seen to need an additional temperature component to fit its SED -- L1521F. This core has recently been reported to contain a protostar, and we confirm this assertion. However, no other core was detected at 24\mum\, and no other core required a hotter dust component to fit its SED. Hence we deduce that the number of misclassified pre-stellar cores is probably small, and the resultant change in the estimated lifetime of the pre-stellar core stage is $<$5 per cent.

\section*{Acknowledgements}

This work is based on observations made with the {\it Spitzer Space Telescope}, which is operated by the Jet Propulsion Laboratory, California Institute of Technology under a contract with NASA. JMK is grateful to the \spitzer\ help desk for their discussions on detector saturation. JMK acknowledges PPARC post-doctoral support at Cardiff University. DWT was on sabbatical at the Observatoire de Bordeaux and CEA Saclay whilst carrying out this work and gratefully acknowledges the hospitality accorded to him there. We thank the anonymous referee for useful comments that have improved this paper.


\begin{thebibliography}{39}
\expandafter\ifx\csname natexlab\endcsname\relax\def\natexlab#1{#1}\fi

\bibitem[{{Alves} {et~al.}(2001){Alves}, {Lada}, \& {Lada}}]{2001alves}
{Alves} J.~F., {Lada} C.~J., {Lada} E.~A., 2001, \nat, 409, 159

\bibitem[\protect\citeauthoryear{Andr{\'e}, Motte, \& 
Bacmann}{1999}]{1999amb} Andr{\'e} P., Motte F., Bacmann A., 
1999, \apj, 513, L57 

\bibitem[\protect\citeauthoryear{Andre, Ward-Thompson, \& 
Barsony}{1993}]{1993awb} Andre P., Ward-Thompson D., Barsony 
M., 1993, ApJ, 406, 122 

\bibitem[{{Andre} {et~al.}(1996){Andre}, {Ward-Thompson}, \& {Motte}}]{paper2}
{Andr{\'e}} P., {Ward-Thompson} D., {Motte} F., 1996, \aap, 314, 625, Paper II

\bibitem[{{Bacmann} {et~al.}(2000){Bacmann}, {Andr{\'e}}, {Puget}, {Abergel},
  {Bontemps}, \& {Ward-Thompson}}]{2000bacmann}
{Bacmann} A., {Andr{\'e}} P., {Puget} J.~L., {Abergel} A., {Bontemps} S.,
  {Ward-Thompson} D., 2000, \aap, 361, 555

\bibitem[{{Beichman} {et~al.}(1986){Beichman}, {Myers}, {Emerson}, {Harris},
  {Mathieu}, {Benson}, \& {Jennings}}]{1986beichman}
{Beichman} C.~A., {Myers} P.~C., {Emerson} J.~P., {Harris} S., {Mathieu} R.,
  {Benson} P.~J., {Jennings} R.~E., 1986, \apj, 307, 337

\bibitem[{{Benson} \& {Myers}(1989)}]{1989bm}
{Benson} P.~J., {Myers} P.~C., 1989, \apjs, 71, 89

\bibitem[Bergin et al.(2006)]{2006bergin} Bergin, E.~A., Maret, 
S., van der Tak, F.~F.~S., Alves, J., Carmody, S.~M., \& Lada, C.~J.\ 2006, 
\apj, 645, 369 

\bibitem[{{Bonnor}(1956)}]{1956bonnor}
{Bonnor} W.~B., 1956, \mnras, 116, 351

\bibitem[\protect\citeauthoryear{Bontemps, Ward-Thompson, \& 
Andre}{1996}]{1996bwa} Bontemps S., Ward-Thompson D., Andre P., 
1996, A\&A, 314, 477 

\bibitem[Bourke et al.(2006)]{2006bourke} Bourke, T.~L., et al.\ 
2006, \apj, 649, L37 

\bibitem[{{Cameron}(1976)}]{1976kuiper}
{Cameron} R.~M., 1976, \skytel, 52, 327

\bibitem[Caselli et al.(2002)]{2002n2hp} Caselli, P., Benson, 
P.~J., Myers, P.~C., \& Tafalla, M.\ 2002, \apj, 572, 238 

\bibitem[Caselli et al.(2002)]{2002caselli} Caselli, P., Walmsley, 
C.~M., Zucconi, A., Tafalla, M., Dore, L., \& Myers, P.~C.\ 2002, \apj, 
565, 331 

\bibitem[{{Codella} {et~al.}(1997){Codella}, {Welser}, {Henkel}, {Benson}, \&
  {Myers}}]{1997codella}
{Codella} C., {Welser} R., {Henkel} C., {Benson} P.~J., {Myers} P.~C., 1997,
  \aap, 324, 203

\bibitem[{{Crapsi} {et~al.}(2005{\natexlab{a}}){Crapsi}, {Caselli}, {Walmsley},
  {Myers}, {Tafalla}, {Lee}, \& {Bourke}}]{2005evo}
{Crapsi} A., {Caselli} P., {Walmsley} C.~M., {Myers} P.~C., {Tafalla} M., {Lee}
  C.~W., {Bourke} T.~L., 2005{\natexlab{a}}, \apj, 619, 379

\bibitem[{{Crapsi} {et~al.}(2005{\natexlab{b}}){Crapsi}, {Devries}, {Huard},
  {Lee}, {Myers}, {Ridge}, {Bourke}, {Evans}, {J{\o}rgensen}, {Kauffmann},
  {Lee}, {Shirley}, \& {Young}}]{2005crapsi}
{Crapsi} A., et al, 2005{\natexlab{b}}, \aap, 439, 1023

\bibitem[{Currie \& Berry(2004)}]{2004kappa}
Currie M.~J., Berry D.~S., 2004, Starlink User Note 95.19, CCLRC

\bibitem[Dunham et al.(2006)]{2006dunham} Dunham, M.~M., et al.\ 
2006, \apj, in press

\bibitem[{{Ebert}(1955)}]{1955ebert}
{Ebert} R., 1955, Zeitschrift fur Astrophysik, 37, 217

\bibitem[Evans et al.(2001)]{evans-cores2} Evans, N.~J., II, 
Rawlings, J.~M.~C., Shirley, Y.~L., \& Mundy, L.~G.\ 2001, \apj, 557, 193 

\bibitem[{{Evans} {et~al.}(2003){Evans}, {Allen}, {Blake}, {Boogert}, {Bourke},
  {Harvey}, {Kessler}, {Koerner}, {Lee}, {Mundy}, {Myers}, {Padgett},
  {Pontoppidan}, {Sargent}, {Stapelfeldt}, {van Dishoeck}, {Young}, \&
  {Young}}]{2003c2d}
{Evans} N.~J., et al, 2003, \pasp, 115, 965

\bibitem[Goodwin et al.(2006)]{2006goodwin} Goodwin, S.~P., Kroupa, 
P., Goodman, A., \& Burkert, A.\ 2006, in Protostars and Planets V, in press 

\bibitem[Goodwin et al.(2004)]{2004goodwin} Goodwin, S.~P., 
Whitworth, A.~P., \& Ward-Thompson, D.\ 2004, \aap, 414, 633 

\bibitem[{{Gordon} {et~al.}(2005){Gordon}, {Rieke}, {Engelbracht}, {Muzerolle},
  {Stansberry}, {Misselt}, {Morrison}, {Cadien}, {Young}, {Dole}, {Kelly},
  {Alonso-Herrero}, {Egami}, {Su}, {Papovich}, {Smith}, {Hines}, {Rieke},
  {Blaylock}, {P{\'e}rez-Gonz{\'a}lez}, {Le Floc'h}, {Hinz}, {Latter},
  {Hesselroth}, {Frayer}, {Noriega-Crespo}, {Masci}, {Padgett}, {Smylie}, \&
  {Haegel}}]{2005mips}
{Gordon} K.~D., et al, {Haegel} N.~M., 2005, \pasp, 117, 503

\bibitem[{{Guieu} {et~al.}(2006){Guieu}, {Dougados}, {Monin}, {Magnier}, \&
  {Mart{\'{\i}}n}}]{2006gdmmm}
{Guieu} S., {Dougados} C., {Monin} J.-L., {Magnier} E., {Mart{\'{\i}}n} E.~L.,
  2006, \aap, 446, 485

\bibitem[{{Hirota} {et~al.}(2002){Hirota}, {Ito}, \& {Yamamoto}}]{2002hiy}
{Hirota} T., {Ito} T., {Yamamoto} S., 2002, \apj, 565, 359

\bibitem[{{Jessop} \& {Ward-Thompson}(2001)}]{paper4}
{Jessop} N.~E., {Ward-Thompson} D., 2001, \mnras, 323, 1025, Paper IV

\bibitem[\protect\citeauthoryear{Kauffmann et al.}{2005}]{2005kauffmann} 
{Kauffmann} J., et al, 2005, AN, 326, 878 

\bibitem[{{Kessler} {et~al.}(1996){Kessler}, {Steinz}, {Anderegg}, {Clavel},
  {Drechsel}, {Estaria}, {Faelker}, {Riedinger}, {Robson}, {Taylor}, \&
  {Ximenez de Ferran}}]{1996iso}
{Kessler} M.~F., et al, 1996, \aap, 315, L27

\bibitem[{{Kirk} {et~al.}(2005){Kirk}, {Ward-Thompson}, \&
  {Andr{\'e}}}]{paper6}
{Kirk} J.~M., {Ward-Thompson} D., {Andr{\'e}} P., 2005, \mnras, 360, 1506,
  Paper VI

\bibitem[{{Kirk} {et~al.}(2006){Kirk}, {Ward-Thompson}, \&
  {Crutcher}}]{2006kwc}
{Kirk} J.~M., {Ward-Thompson} D., {Crutcher} R.~M., 2006, \mnras, 369, 1445

\bibitem[Larson(1981)]{larson81} Larson, R.~B.\ 1981, \mnras, 
194, 809 

\bibitem[Launhardt et al.(1997)]{1997launhardt} Launhardt, R., 
Ward-Thompson, D., \& Henning, T.\ 1997, \mnras, 288, L45 

\bibitem[Lee \& Myers(1999)]{1999lee} Lee, C.~W., \& Myers, 
P.~C.\ 1999, \apjs, 123, 233 

\bibitem[{LEOPARD(2005)}]{2005leopard}
LEOPARD, 2005, Leopard User's Guide, v4.1 edn. SPITZER Science Center

\bibitem[{{Makovoz} \& {Khan}(2005)}]{2005mopex}
{Makovoz} D., {Khan} I., 2005, in Astronomical Society of the Pacific
  Conference Series, {Shopbell} P., {Britton} M., {Ebert} R., eds., p.~81

\bibitem[{MIPS(2006)}]{2006mips}
MIPS, 2006, Multiband Imagine Photometer for SPITZER (MIPS) Data Handbook,
  v3.2.1 edn. SPITZER Science Center

\bibitem[Motte et al.(1998)]{1998man} Motte, F., Andre, P., \& 
Neri, R.\ 1998, \aap, 336, 150 

\bibitem[Motte et al.(2001)]{2001motte} Motte, F., Andr{\'e}, P., 
Ward-Thompson, D., \& Bontemps, S.\ 2001, \aap, 372, L41 

\bibitem[Myers et al.(1991)]{1991myers} Myers, P.~C., Fuller, 
G.~A., Goodman, A.~A., \& Benson, P.~J.\ 1991, \apj, 376, 561 

\bibitem[Nutter \& Ward-Thompson(2006)]{2006nw} Nutter, 
D., \& Ward-Thompson, D.\ 2006, \mnras, in press

\bibitem[{{Nutter} {et~al.}(2006){Nutter}, {Ward-Thompson}, \&
  {Andr{\'e}}}]{2006nwa}
{Nutter} D., {Ward-Thompson} D., {Andr{\'e}} P., 2006, \mnras, 368, 1833

\bibitem[\protect\citeauthoryear{Onishi, Mizuno, \& Fukui}{1999}]{1999onishi} 
Onishi T., Mizuno A., Fukui Y., 1999, PASJ, 51, 257 

\bibitem[Onishi et al.(1998)]{1998onishi} Onishi, T., Mizuno, A., 
Kawamura, A., Ogawa, H., \& Fukui, Y.\ 1998, \apj, 502, 296 

\bibitem[{{Onishi} {et~al.}(2002){Onishi}, {Mizuno}, {Kawamura}, {Tachihara},
  \& {Fukui}}]{2002omkt}
{Onishi} T., {Mizuno} A., {Kawamura} A., {Tachihara} K., {Fukui} Y., 2002,
  \apj, 575, 950
  
\bibitem[{{Rieke} {et~al.}(2004){Rieke}, {Young}, {Engelbracht}, {Kelly},
  {Low}, {Haller}, {Beeman}, {Gordon}, {Stansberry}, {Misselt}, {Cadien},
  {Morrison}, {Rivlis}, {Latter}, {Noriega-Crespo}, {Padgett}, {Stapelfeldt},
  {Hines}, {Egami}, {Muzerolle}, {Alonso-Herrero}, {Blaylock}, {Dole}, {Hinz},
  {Le Floc'h}, {Papovich}, {P{\'e}rez-Gonz{\'a}lez}, {Smith}, {Su}, {Bennett},
  {Frayer}, {Henderson}, {Lu}, {Masci}, {Pesenson}, {Rebull}, {Rho}, {Keene},
  {Stolovy}, {Wachter}, {Wheaton}, {Werner}, \& {Richards}}]{2004mips}
{Rieke} G.~H., et al, 2004, \apjs, 154, 25

\bibitem[{{Shirley} {et~al.}(2000){Shirley}, {Evans}, {Rawlings}, \&
  {Gregersen}}]{2000serg}
{Shirley} Y.~L., {Evans} N.~J., {Rawlings} J.~M.~C., {Gregersen} E.~M., 2000,
  \apjs, 131, 249

\bibitem[Stamatellos \& Whitworth(2003)]{2003sw} Stamatellos, 
D., \& Whitworth, A.~P.\ 2003, \aap, 407, 941 

\bibitem[{{Stamatellos} {et~al.}(2005){Stamatellos}, {Whitworth}, {Boyd}, \&
  {Goodwin}}]{2005swbg}
{Stamatellos} D., {Whitworth} A.~P., {Boyd} D.~F.~A., {Goodwin} S.~P., 2005,
  \aap, 439, 159

\bibitem[{{Steinacker} {et~al.}(2005){Steinacker}, {Bacmann}, {Henning},
  {Klessen}, \& {Stickel}}]{2005steinacker}
{Steinacker} J., {Bacmann} A., {Henning} T., {Klessen} R., {Stickel} M., 2005,
  \aap, 434, 167

\bibitem[\protect\citeauthoryear{Straizys et al.}{1992}]{1992straizys} 
Straizys V., Cernis K., Kazlauskas A., Meistas E., 1992, BaltA, 1, 149 

\bibitem[{{Tafalla} \& {Santiago}(2004)}]{2004ts}
{Tafalla} M., {Santiago} J., 2004, \aap, 414, L53

\bibitem[{{Visser} {et~al.}(2001){Visser}, {Richer}, \& {Chandler}}]{2001vrc}
{Visser} A.~E., {Richer} J.~S., {Chandler} C.~J., 2001, \mnras, 323, 257

\bibitem[{{Ward-Thompson} {et~al.}(2005){Ward-Thompson}, {Ade}, {Araujo},
  {Coulson}, {Cox}, {Davis}, {Evans}, {Griffin}, {Gear}, {Hargrave},
  {Hargreaves}, {Hayton}, {Kiernan}, {Leeks}, {Mauskopf}, {Naylor}, {Potter},
  {Rinehart}, {Sudiwala}, {Tucker}, {Walker}, \& {Watkin}}]{2005thumper}
{Ward-Thompson} D., et al, 2005, \mnras, 364, 843

\bibitem[{{Ward-Thompson} {et~al.}(2006){Ward-Thompson}, {Andre}, {Crutcher},
  {Johnstone}, {Onishi}, \& {Wilson}}]{2006wtppv}
{Ward-Thompson} D., {Andr{\'e}} P., {Crutcher} R., {Johnstone} D., {Onishi} T.,
  {Wilson} C., 2006, in Protostars and Planets V, in press

\bibitem[{{Ward-Thompson} {et~al.}(2002){Ward-Thompson}, {Andr{\'e}}, \&
  {Kirk}}]{paper5}
{Ward-Thompson} D., {Andr{\'e}} P., {Kirk} J.~M., 2002, \mnras, 329, 257, Paper
  V

\bibitem[\protect\citeauthoryear{Ward-Thompson et 
al.}{1995}]{1995wckab} Ward-Thompson D., Chini R., Krugel E., 
Andre P., Bontemps S., 1995, MNRAS, 274, 1219 

\bibitem[{{Ward-Thompson} {et~al.}(1999){Ward-Thompson}, {Motte}, \&
  {Andre}}]{paper3}
{Ward-Thompson} D., {Motte} F., {Andr{\'e}} P., 1999, \mnras, 305, 143, Paper III

\bibitem[{{Ward-Thompson} {et~al.}(1994){Ward-Thompson}, {Scott}, {Hills}, \&
  {Andre}}]{paper1}
{Ward-Thompson} D., {Scott} P.~F., {Hills} R.~E., {Andr{\'e}} P., 1994, \mnras,
  268, 276, Paper I

\bibitem[{{Werner} {et~al.}(2004){Werner}, {Roellig}, {Low}, {Rieke}, {Rieke},
  {Hoffmann}, {Young}, {Houck}, {Brandl}, {Fazio}, {Hora}, {Gehrz}, {Helou},
  {Soifer}, {Stauffer}, {Keene}, {Eisenhardt}, {Gallagher}, {Gautier}, {Irace},
  {Lawrence}, {Simmons}, {Van Cleve}, {Jura}, {Wright}, \&
  {Cruikshank}}]{2004spitzer}
{Werner} M.~W., et al, 2004, \apjs, 154, 1

\bibitem[Young et al.(2004)]{2004young} Young, C.~H., et al.\ 
2004, \apjs, 154, 396 

\bibitem[Young et al.(2006)]{2006young} Young, C.~H., et al.\ 
2006, in press 

\end{thebibliography}
\end{document}